%% file: main.tex
\documentclass[acmtog,nonacm]{acmart}

\usepackage{listings}

\lstdefinelanguage{json}{
  basicstyle=\normalfont\ttfamily,
  numbers=left,
  numberstyle=\tiny,
  stepnumber=1,
  numbersep=5pt,
  showstringspaces=false,
  breaklines=true,
  frame=lines,
  backgroundcolor=\color{white},
  literate={"}{{\texttt{\char34}}}1
           {,}{{\texttt{,}}}1
           {:}{{\texttt{:}}}1
           {[}{{\texttt{[}}}1
           {]}{{\texttt{]}}}1
           {\{}{{\texttt{\{}}}1
           {\}}{{\texttt{\}}}}1
}

\AtBeginDocument{
  }

\begin{document}

\title{HairGPT: Strand-as-Language Autoregressive Modeling for Realistic 3D Hairstyle Synthesis}

\author{Haimin Luo}
\affiliation{
  \institution{ShanghaiTech University}
  \city{Shanghai}
  \country{China}}
\email{luohm@shanghaitech.edu.cn}

\author{Min Ouyang}
\affiliation{
  \institution{ShanghaiTech University and Deemos Technology Co., Ltd.}
  \city{Shanghai}
  \country{China}}
\email{ouyangmin2022@shanghaitech.edu.cn}

\author{Lan Xu}
\affiliation{
  \institution{ShanghaiTech University}
  \city{Shanghai}
  \country{China}}
\email{xulan1@shanghaitech.edu.cn}

\author{Jingyi Yu}
\affiliation{
  \institution{ShanghaiTech University}
  \city{Shanghai}
  \country{China}
}
\email{yujingyi@shanghaitech.edu.cn}

\renewcommand\shortauthors{Luo, H. et al}

\begin{abstract}

Hair is a rich medium of visual and cultural expression, yet its digital modeling remains challenging due to the duality of fluidity and structure. Many existing generative approaches rely primarily on continuous diffusion fields, which entangle global topology with local texture and obscure the semantic and structural organization of hairstyles. To address this, we propose HairGPT, a strand-centric framework that treats strands as generative primitives and formulates realistic 3D hairstyle synthesis as a dual-decoupled autoregressive sequence modeling problem. Our method applies spatial decoupling across semantic scalp regions and structural decoupling along a hierarchical strand representation, progressing from global layout to fine-grained style. We further introduce a geometric tokenizer and region-aware semantic annotations to guide strand-level generation, enabling compositional editing, synthesis of rare and complex hairstyles, and adaptation to stylized domains. By aligning generative modeling with the workflow of digital grooming, HairGPT turns hair generation from opaque texture synthesis into a structured and semantically controllable authoring process, supporting robust semantic conditioning and high-fidelity results across realistic and stylized domains.

\end{abstract}



\begin{teaserfigure}
  \centering
\includegraphics[width=1\linewidth]{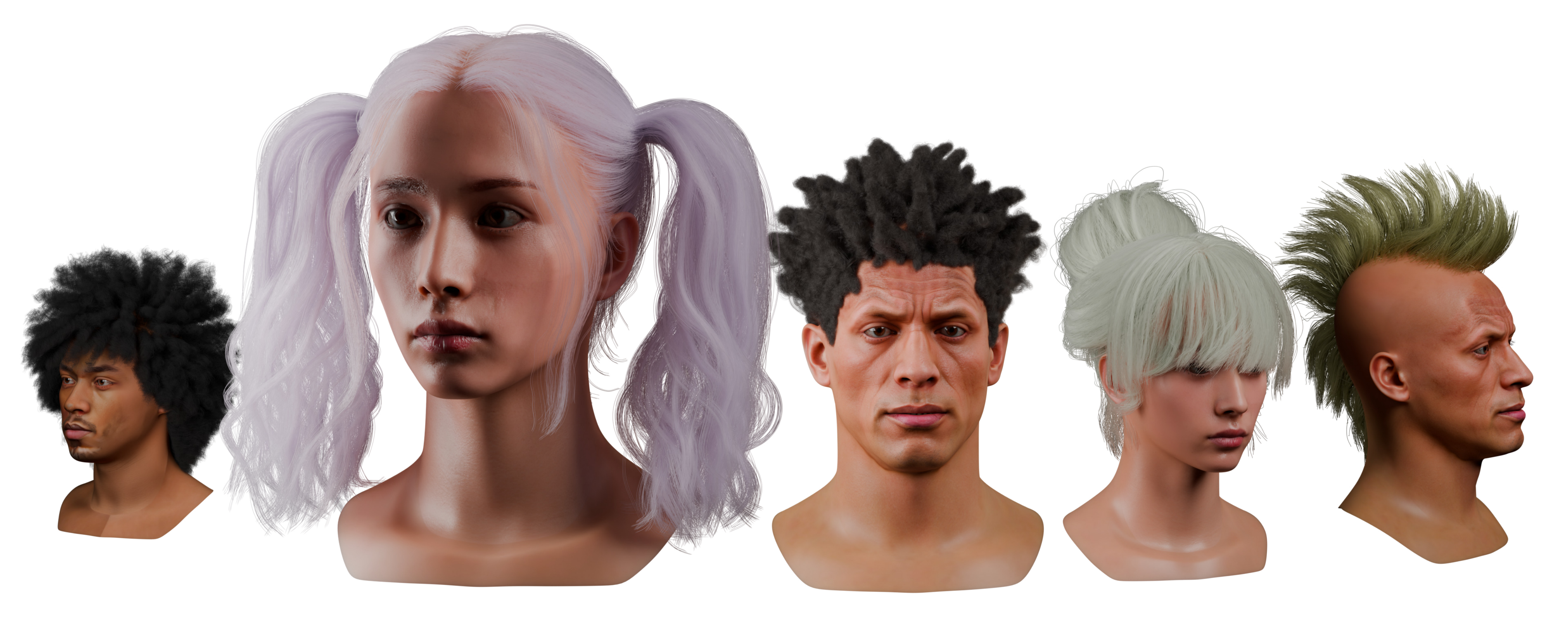}
  \caption{We introduce ``HairGPT'', a unified autoregressive framework for realistic 3D hairstyle synthesis. HairGPT uses individual strands as the fundamental generative units and formulates hair generation as structured sequence modeling over strands. By integrating image, text, and hair modalities, HairGPT supports robust cross-modal conditioning. This unified formulation enables the generation of high-frequency hair details and complex topological structures, translating linguistic instructions and visual cues into high-fidelity 3D strands.}
  \Description{Overview teaser showing HairGPT generating realistic 3D hairstyles from image and text conditions through a strand-based autoregressive framework.}
  \label{fig:teaser}
\end{teaserfigure}
\maketitle

\input{sec/1_intro}
\input{sec/2_related_work}

\input{sec/3_0_overview}
\input{sec/3_1_hair_rep}
\input{sec/3_2_hair_model}
\input{sec/4_results}

\input{sec/5_conclusion}

\begin{acks}
This work was supported in part by the National Natural Science Foundation of China under Grant W2431046, Central Guided Local Science and Technology Foundation of China YDZX20253100001001, and by MoE Key Lab of Intelligent Perception and Human-Machine Collaboration (ShanghaiTech University), the Shanghai Frontiers Science Center of Human-centered Artificial Intelligence. This work was also supported by the HPC Platform of ShanghaiTech University.

The authors would also like to thank Heng'an Zhou from ShanghaiTech University for his assistance with the supplementary video, and Zijun Zhao from Deemos Technology Co., Ltd.\ for helping to process part of the raw hairstyle data. 
\end{acks}

\bibliographystyle{ACM-Reference-Format}
\bibliography{main}

\clearpage
\appendix
\input{sec/supplementary}

\end{document}

%% file: sec/1_intro.tex
\section{Introduction}

In the physical world, hair is far more than a biological filament; it is the ``silken alphabet'' of human identity. From the rebellious punk mohawk to the ceremonial braided bun, hair serves as a dynamic manifesto of culture, gender, and personality. This significance is not merely visual: hair is one of the most broadly shared media of human expression, woven into everyday life through acts of styling, care, and self-presentation. Its richness lies not only in appearance, but also in deliberate authorship and subtle structural design, through which strands are arranged into meaningful and culturally legible forms. Digital hair creation, therefore, should make this authorship equally accessible, translating personal intent into structured, controllable 3D geometry.

Realizing this goal, however, is difficult precisely because hair is not generic geometry to be reproduced, but an intentionally organized structure to be modeled and controlled. Its complexity lies in a paradox: hair is fluid yet structured, chaotic yet groomed. Capturing this duality digitally is not merely a geometric challenge; it is equally an artistic one. Hairstyles are not perceived as arbitrary collections of fibers, but as organized forms governed by anatomical regions and aesthetic intent. Accordingly, generative hair creation should not be framed as holistic synthesis over an undifferentiated signal, but as structured authoring over hierarchically controlled primitives.

Most existing generative approaches~\cite{sklyarova2023haar, rosu2025difflocks}, however, cast hair generation as a 2D texture synthesis problem. By projecting 3D geometry into latent texture maps parameterized by the scalp's UV coordinates, these methods leverage 2D diffusion over continuous geometric fields. While visually compelling, this paradigm fundamentally misaligns with the structured nature of hair. Flattening a 3D composition into a single monolithic latent field inherently entangles global topology with local texture. Consequently, compositional editing---such as modifying curl tightness without destroying the global silhouette, or adjusting bangs without disturbing the crown---becomes brittle or infeasible. This limitation prevents predictable, region-specific control for high-fidelity editing. More broadly, it leaves a profound semantic gap between how human creators intuitively describe hairstyles and how the model internally represents them.

We revisit the strand as the fundamental generative unit, casting hair generation as a structured sequence modeling problem. Rather than denoising an entire hairstyle holistically, we generate hair autoregressively as an ordered assembly of strand groups. To make this tractable for high-fidelity assets, we first condense dense raw hair models into sparse sets of representative guide strands, providing a compact yet expressive scaffold. By mapping these spatial strands into a discrete geometric vocabulary, autoregressive transformers can model hair as a sequence in which each structural decision is conditioned on previously generated geometry. In this view, realistic hair is no longer a monolithic field to be synthesized all at once, but a structured authoring process that progressively organizes strands into coherent forms.

Building on this strand-centric philosophy, we introduce HairGPT, a framework driven by Dual Decoupling along two orthogonal axes. First, we apply spatial decoupling by partitioning the scalp into semantically distinct regions, enabling localized synthesis and targeted editing. Second, we apply structural decoupling by decomposing strand geometry into a hierarchy of generative stages, progressing from global density and layout, to coarse strand shape, and finally to fine-grained style residuals.

Crucially, generation is not performed over a single monolithic sequence. Instead, we introduce a mode-specific autoregressive strategy. The model is trained and sampled under distinct layout, coarse, and style modes. Each mode constructs its own sequence over strand units, avoiding unnecessary coupling across stages and mitigating error accumulation. By decomposing geometry into this hierarchy, HairGPT internalizes the structured logic of professional grooming: it establishes global organization before refining local detail, allowing the model to master one level of abstraction at a time without requiring users to manually author these complex topological steps.

This structural foundation supports more accessible hair creation by exposing controllable abstractions that align with high-level human intent. To connect this structural representation with human-readable control, we introduce a region-aware text annotation pipeline. By fine-tuning a pretrained vision-language model, we extract both global hairstyle descriptions and region-specific local attributes aligned with our scalp partition. Injecting these linguistic tokens into the autoregressive sequence provides semantic guidance for the generative process, allowing intuitive text-driven control without sacrificing structural consistency.

Together, these design choices transform hair generation from an opaque texture synthesis problem into a transparent, compositional modeling process. By aligning representation, training, and inference with the organizational logic of strands and regions, HairGPT enables controllable synthesis, editing, and robust generation of complex hairstyles.

In summary, our contributions are:
\begin{itemize}
\item We introduce a \textbf{strand-as-language} paradigm for realistic 3D hair generation, reformulating the problem as a dual-decoupled autoregressive process in which strands serve as the fundamental generative units.

\item We propose a novel geometric tokenization scheme for guide strands, based on multi-head product quantization, which efficiently encodes complex topology and high-frequency style details into a compact discrete vocabulary.

\item We develop a hierarchical strand-language construction together with a multi-stage training strategy, organizing generation into region-aware and stage-specific sequences for improved structural coherence and stable optimization.

\item We show that HairGPT supports robust semantic conditioning and compositional editing, enabling high-fidelity generation of rare and complex hairstyles, as well as effective adaptation to stylized domains.
\end{itemize}

%% file: sec/2_related_work.tex
\section{Related Work} \label{sec:related_work}
\subsection{3D Hair Representation}

\begin{figure*}[t]
	\centering
	\includegraphics[width=\linewidth]{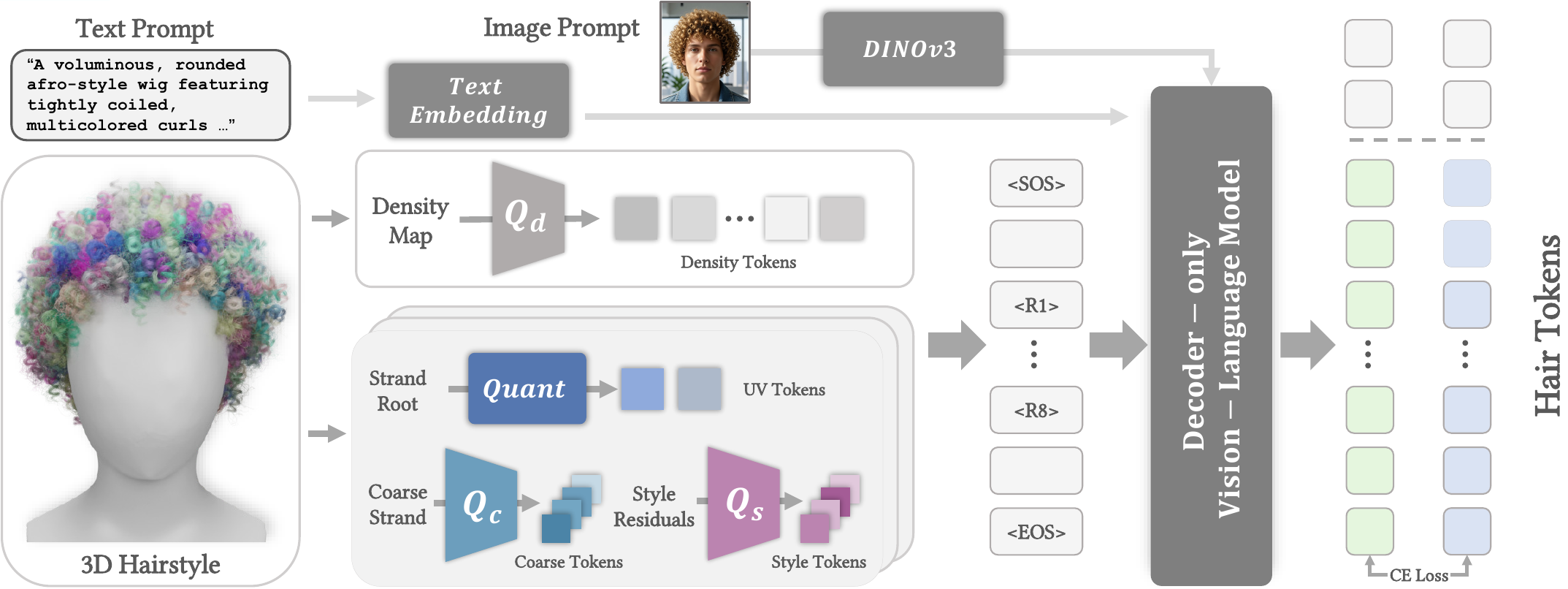}
	\caption{\textbf{HairGPT Overview.} The 3D hairstyle geometry is decomposed into a global density map (quantized by tokenizer $Q_d$) and local strand features. Specifically, strand roots are encoded into two UV tokens. The strand geometry is decoupled into coarse shape and style residuals, which are further discretized into four tokens by tokenizers $Q_c$ and $Q_s$. These geometric codes are assembled into a hierarchical sequence, which is processed by a decoder-only Transformer. After concatenation with text and image embeddings, the model autoregressively predicts the target hair tokens and is supervised via cross-entropy loss.
  }
  \Description{Pipeline diagram showing decomposition of 3D hair into density, root, coarse, and style tokens, followed by hierarchical autoregressive prediction with image and text conditioning.}
	\label{fig:pipeline}
\end{figure*}

\paragraph{Traditional Geometric Modeling.}
Early attempts to model hair focused primarily on explicit parametric geometric representations, such as parametric surfaces~\cite{koh2000real, liang2003enhanced, noble2004modelling}, wisps and generalized cylinders~\cite{chen1999system, yang2000cluster, xu2001v, patrick2004modelling, choe2005statistical, kim2002interactive, wang2004hair}, and hair meshes~\cite{yuksel2009hair}. However, these models typically require significant manual labor to create realistic hairstyles. 
Subsequent image-based hair modeling efforts~\cite{19981351, grabli2002image, paris2004capture} sought to synthesize strands according to a 3D hair flow volume in a heuristic manner. Other works~\cite{wei2005modeling, luo2012multi, luo2013wide, paris2008hair, lightinghair} employed high-end acquisition systems to capture accurate 3D hair orientation fields, with hair primitives such as wisps and strands derived through subsequent growing procedures. Such sophisticated capture systems prevent these methods from being universally accessible.

\paragraph{Volumetric/Neural Representations.} 
More recent works have introduced neural rendering techniques based on volumetric representations, such as neural orientation fields~\citep{wu2022neuralhdhair, kuang2022deepmvshair, sklyarova2023neural}, volumetric points~\citep{Wang2020NOPC}, and neural radiance fields~\citep{mildenhall2021nerf, convnerf, Artemis, wang2022hvh, wang2023neuwigs, wu2024monohair, zhou2024groomcap}. These approaches primarily prioritize appearance and rendering, though the resulting geometry is often constrained by the low resolution of the underlying representations. GaussianHair~\cite{luo2024gaussianhair} and subsequent works~\citep{zhou2024groomcap, zakharov2024human, zheng2025groomlight, Pan_2025_BMVC} reformulate individual hair strands as sequences of cylindrical Gaussians, leveraging the popular 3DGS~\cite{kerbl3Dgaussians} framework to model hair from images. These methods achieve high-fidelity strand reconstruction and facilitate efficient real-time rendering. However, they remain fundamentally limited by time-consuming per-scene optimization processes, which lack the flexibility and speed required for interactive, prompt-based hairstyle synthesis.

\subsection{Generative and Reconstructive Hair Modeling}

\paragraph{Sparse-view Reconstruction.} 
Reconstructing 3D hair from sparse-view images is an ill-posed task that requires strong generative priors. Early data-driven methods~\cite{hu2014robust, yu2014hybrid, zhang2017data, hu2015single} relied on synthetic hairstyle databases containing complete strand geometries. These methods search the database to identify hairstyles that best match the observed sparse views.
More recent deep learning-based works~\cite{wu2022neuralhdhair, kuang2022deepmvshair, zhou2018hairnet, saito20183d, yang2019dynamic, rosu2022neural, sklyarova2023neural, autohair, 4viewhair, zheng2023hairstep, shen2023CT2Hair, takimoto2024drhair, wu2024monohair} instead learn shape priors from synthetic hair datasets. In these frameworks, an intermediate 3D hair orientation volume is inferred from sparse image inputs to guide strand-growing algorithms. However, such works tend to be constrained by the limited diversity of hair datasets.

\paragraph{Generative Synthesis.} 
Driven by rapid advancements in 3D geometry synthesis, realistic hairstyle generation has evolved from VAE-based latent exploration~\cite{zhou2023groomgen} to diffusion-based~\cite{rosu2025difflocks} models. \textit{HAAR}~\cite{sklyarova2023haar} established a text-to-strand generation framework, while TANGLED~\cite{long2025tangled} leverages a multi-view line-art conditional diffusion approach. DiffLocks~\cite{rosu2025difflocks} proposed a scalp-texture diffusion model capable of handling diverse textures, including highly curled and afro-style hair. However, these works often treat hair as an entangled texture, overlooking the anatomical partitioning and structural hierarchy that a human stylist intuitively employs to manage geometric complexity.
Alternatively, autoregressive (AR) models treat hair as a sequential ``hair language.'' CHARM~\cite{he2025charm} pioneered this by modeling anime hair cards as sequences of control points. Our HairGPT pushes this paradigm further by treating realistic 3D hair geometry as a native linguistic modality, unifying reconstruction and creative synthesis within a single vision-language transformer.

\subsection{Vision-Language Models for 3D Generation}
The integration of Large Multimodal Models (LMMs) has popularized the ``Geometry-as-Language'' paradigm. Pioneering works such as PolyGen~\cite{nash2020polygen} and MeshGPT~\cite{Siddiqui2024MeshGPT} established the foundation for mesh synthesis via discrete token prediction. Building upon these, recent frameworks like Argus~\cite{argus2025} and MeshAnythingV2~\cite{chen2024meshanything} demonstrate the potential of transformers to synthesize general meshes with high topological fidelity.
In the hair domain, Hairmony~\cite{Meishvili2024Hairmony} established a rigorous multidisciplinary taxonomy grounded in anthropology, hair science, and professional grooming, providing a comprehensive classification system for complex hairstyle attributes. Our HairGPT bridges the gap between this taxonomy and geometric synthesis by being the first to operationalize these principles within a native vision-language model.

%% file: sec/3_0_overview.tex
\section{Overview}

At the core of HairGPT lies a dual-decoupled hairstyle representation, arising naturally from our view that hair is not generic geometry to be reproduced, but human-authored structure to be organized and controlled. This representation makes it possible to cast hairstyle synthesis as autoregressive generation over explicit strand units, while translating high-level human intent into structured 3D geometry. 
As illustrated in Fig.~\ref{fig:pipeline}, we first derive a compact structural abstraction from dense 3D hairstyle geometry, consisting of a global scalp density map and a sparse set of guide strands. Each guide strand is then represented in a dual-decoupled form, where its spatial root is separated from its strand-level geometry, and the latter is further decomposed into a low-frequency coarse backbone and a high-frequency style residual. This structured representation, corresponding to the left and middle parts of Fig.~\ref{fig:pipeline}, forms the focus of Sec.~\ref{sec:parameterization}. 

Built upon this representation, HairGPT further discretizes the density map and strand components into tokens, assembles them into region-aware hierarchical sequences, and autoregressively predicts them under joint image-text conditioning using a decoder-only transformer. This generative framework, corresponding to the right part of Fig.~\ref{fig:pipeline}, is presented in Sec.~\ref{sec:hairgpt}. In this way, Sec.~\ref{sec:parameterization} establishes the core hairstyle parameterization, while Sec.~\ref{sec:hairgpt} shows how this parameterization is operationalized within a multimodal autoregressive framework for structured hairstyle synthesis.

%% file: sec/3_1_hair_rep.tex
\section{Dual-Decoupled Hairstyle Parameterization}
\label{sec:parameterization}

We represent hairstyles as collections of individual strands, reflecting a grooming-inspired construction process in which strands are progressively placed and shaped over the scalp.
Based on this view, our goal is to derive a compact strand-based representation from dense raw hair geometry that makes explicit both the global organization of hair over the scalp and the local geometric structure of individual strands.
To this end, we decouple hairstyle modeling along two orthogonal axes: scalp-level spatial organization, which determines where strands are placed, and strand-level geometry, which determines how each strand is shaped from coarse flow to fine detail. This section introduces the resulting dual-decoupled parameterization, together with the guide-strand abstraction and strand decomposition that serve as the structural foundation of HairGPT.

\subsection{The Dual-Decoupled Representation}
\label{sec:representation_concept}

We introduce a dual-decoupled representation that separates scalp-level spatial organization from strand-level geometric structure.
It provides a structured interface for independently modeling strand placement and geometry within a sequential generation process.

\begin{figure}[t]
  \includegraphics[width=1.0\linewidth]{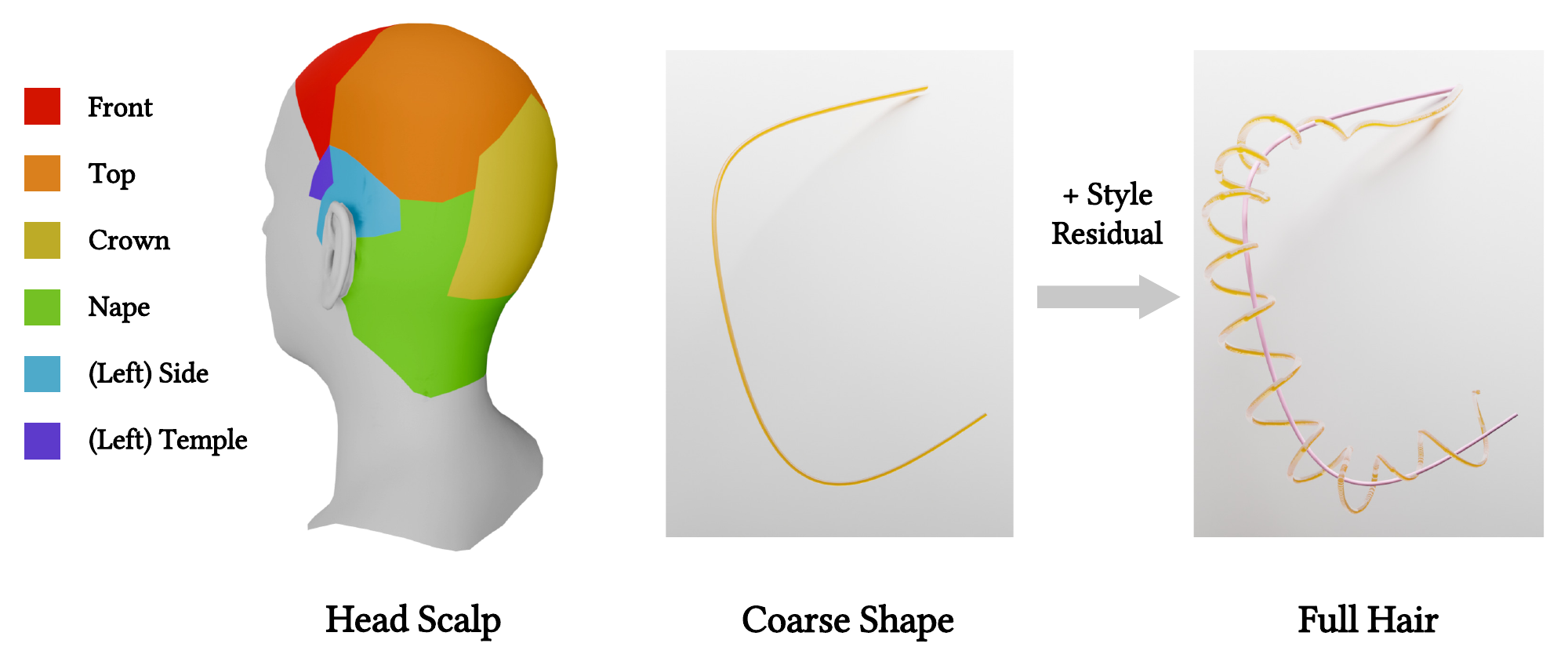}
  \caption{\textbf{Dual-Decoupled Representation.} The scalp is semantically partitioned into eight regions, and each strand is decoupled into a low-frequency coarse backbone and a high-frequency style residual.}
  \Description{Diagram of scalp regions and strand decomposition into root location, coarse backbone, and style residual components.}
  \label{fig:hair_representation}
\end{figure}

\paragraph{Spatial Decoupling via Scalp Partitioning.}
At the macroscopic level, hairstyle generation is formulated as a spatial planning problem on the 2D scalp manifold $\mathcal{S}$.
Following the semantic taxonomy of \textit{Hairmony}~\cite{Meishvili2024Hairmony}, we partition $\mathcal{S}$ into $M=8$ regions
$\mathcal{R} = \{\text{Front}, \text{Top}, \text{Crown}, \text{Nape}, \text{Right/Left~Temple}, \text{Right/Left~Side}\}$ (Fig.~\ref{fig:hair_representation}).
This partitioning provides an explicit spatial prior that enforces locality during strand generation.
In addition, we define a scalp density map $\mathcal{D}: \mathcal{S} \to \mathbb{R}^{+}$ to guide root placement independently of strand geometry.

\paragraph{Structural Decoupling via Strand Hierarchy.}
At the strand level, we model geometry generation itself as a hierarchical process.
The geometry of a strand $\mathbf{s}$ is decomposed into a structured tuple
$\mathbf{s} = (\mathbf{u}, \mathbf{c}, \mathbf{r})$,
reflecting a coarse-to-fine synthesis within each autoregressive step.
The root location $\mathbf{u} \in \mathcal{S}$ anchors the strand to the spatial plan.
The coarse shape $\mathbf{c}$ encodes the low-frequency backbone of the strand,
capturing global flow and topology. The style residual $\mathbf{r}$ adds high-frequency geometric detail (Fig.~\ref{fig:hair_representation}),
yielding the final strand geometry $\mathbf{s} = \mathbf{c} + \mathbf{r}$.
This explicit separation allows the model to reuse fine-scale style patterns across different global strand shapes,
which is essential for learning strand-level stylistic regularities within a sequential generative process.

\subsection{Sparse Guide Strand Extraction}
\label{sec:derivation}

Directly processing a raw hair model $\mathcal{H}$ is impractical due to its scale ($\sim10^5$ strands) and the lack of an inherent strand-level ordering.
We therefore condense the dense geometry into a sparse set of representative \emph{Guide Strands} via clustering, yielding a strand-level sequence suitable for autoregressive modeling.

\paragraph{Frequency-Aware Strand Clustering.}

To reduce sensitivity to high-frequency noise, we perform clustering in a frequency-based feature space.
Since hair strands are open, non-periodic 3D curves, we use the Discrete Cosine Transform (DCT) rather than the Discrete Fourier Transform (DFT): the periodic assumption of the DFT would introduce an artificial root-tip discontinuity, and low-frequency truncation may produce ringing near strand boundaries.
The DCT avoids this issue through its implicit symmetric extension, yielding smoother low-frequency descriptors for strand shape analysis~\cite{NEURIPS2024_DH_hairgen}.

Each raw strand $\mathbf{s}_i \in \mathcal{H}_{\text{raw}}$ is mapped to a compact shape descriptor $\mathbf{z}_i$ using the Discrete Cosine Transform (DCT), after subtracting its root position to remove global translation:
\begin{equation}
    \mathbf{z}_i = \mathcal{T}_{K_{\text{feat}}} \left( \text{DCT}(\mathbf{s}_i - \mathbf{s}_{i,\text{root}}) \right) \in \mathbb{R}^{K_{\text{feat}} \times 3}.
\end{equation}
We retain the first $K_{\text{feat}}=8$ coefficients, which act as a low-pass representation capturing the dominant strand shape.
The strands are then grouped using k-means clustering into $N_{\text{guide}}=512$ clusters by minimizing the standard intra-cluster variance $\sum \| \mathbf{z}_i - \boldsymbol{\mu}_j \|^2$.

\begin{figure}
    \centering
    \includegraphics[width=1.0\linewidth]{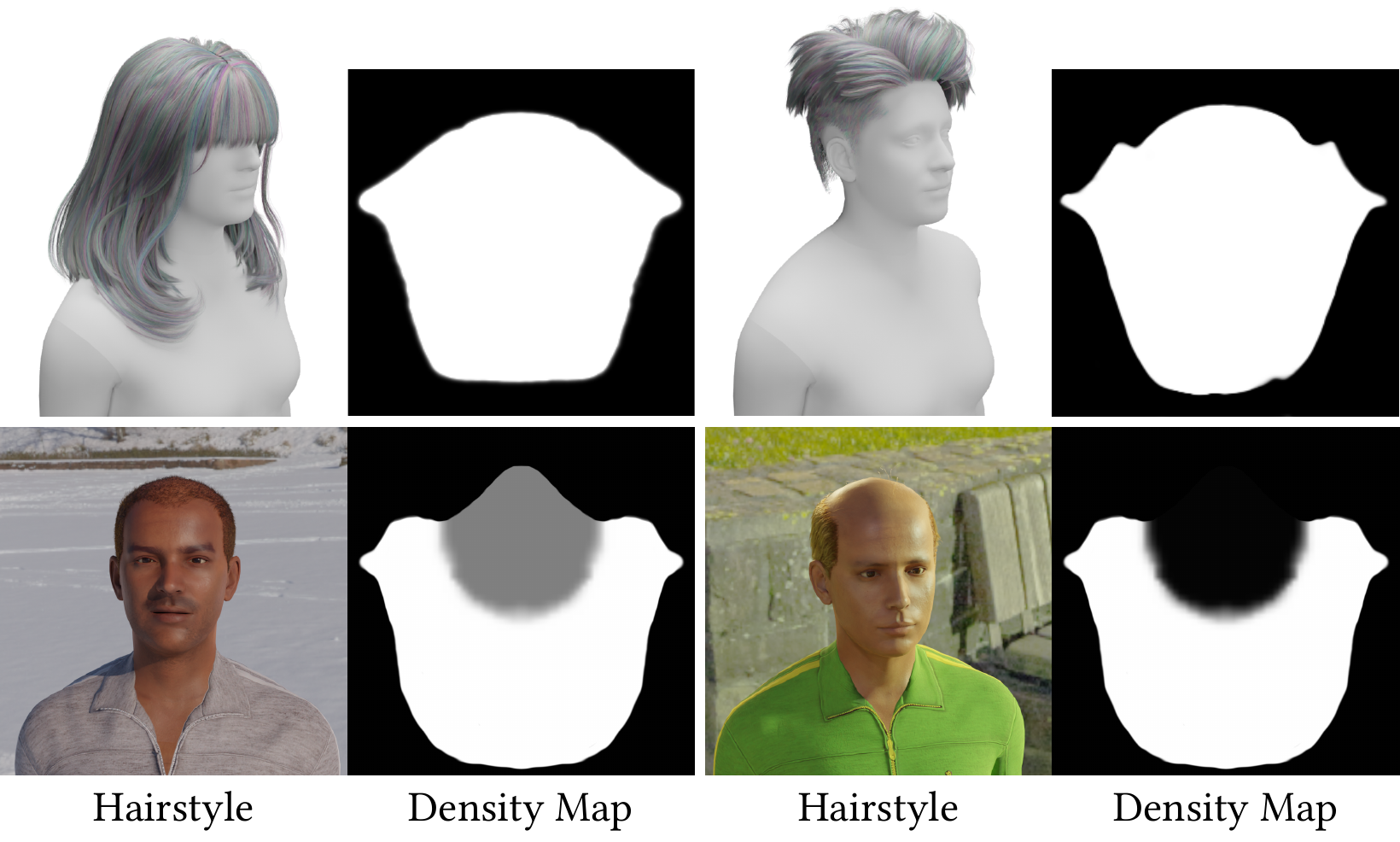}
    \caption{\textbf{Continuous density maps in our dataset.} We show several representative hairstyles together with their corresponding scalp-space density maps.}
    \Description{Examples of hairstyles paired with continuous scalp-space density maps that visualize hair-root distributions.}
    \label{fig:density_visualization}
\end{figure}

For each cluster, the centroid strand is selected as the guide strand, providing a compact geometric proxy for the dense hair volume.
In parallel, we project the root positions of all raw strands onto the scalp UV domain to compute a density map $\mathcal{D}$, which represents the spatial distribution of hair roots over the scalp, similar to DiffLocks~\cite{rosu2025difflocks}. Each continuous value in $\mathcal{D}$ reflects the relative likelihood of a strand root occurring at the corresponding UV location, thereby providing a global prior for strand placement during generation. We visualize representative hairstyles together with their corresponding scalp-space density maps in Fig.~\ref{fig:density_visualization}.

Together, the extracted guide strands provide a sparse, strand-level abstraction suitable for autoregressive modeling.

\subsection{Spectral-Spatial Decomposition}
\label{sec:decomposition}
Based on the extracted guide strands, we further decompose each strand into a coarse geometric backbone and a fine-scale style residual, instantiating the strand-level hierarchy introduced in Sec.~\ref{sec:representation_concept}.

\paragraph{Frequency-Based Coarse Geometry.} 
A guide strand is represented as an ordered point sequence $\mathbf{P} = \{\mathbf{p}_0, \dots, \mathbf{p}_{L-1}\}$.
We operate on segment direction vectors $\mathbf{v}_j = \mathbf{p}_{j+1} - \mathbf{p}_j$ and apply the Discrete Cosine Transform (DCT) to the sequence $\mathbf{V}=\{\mathbf{v}_j\}$, which more directly captures the intrinsic flow of the strand.
Retaining the first $K_{\text{geo}}=4$ coefficients yields filtered directions $\hat{\mathbf{V}}$, which are recovered via inverse DCT:
\begin{equation}
    \hat{\mathbf{V}} = \text{IDCT}\left( \mathcal{T}_{K_{\text{geo}}} \left( \text{DCT}(\mathbf{V}) \right) \right).
\end{equation}
The coarse geometry $\hat{\mathbf{P}}=\{\hat{\mathbf{p}}_j\}$ is reconstructed by cumulative integration: $\hat{\mathbf{p}}_j = \mathbf{p}_0 + \sum_{m=0}^{j-1} \hat{\mathbf{v}}_m$,
yielding a smooth backbone curve that captures global strand flow.

\paragraph{Scaled Style Residuals.}
To encode fine-scale strand details independently of global orientation and physical scale, we define style residuals as normalized deviations from the coarse backbone, expressed in a local coordinate system along the strand.
Notably, although the coarse backbone is extracted via DCT, the residuals are not simply the discarded high-frequency signal; instead, they are defined in a scale-normalized local frame to capture reusable strand-level style patterns.

Specifically, we first introduce a local scale factor $\sigma_j$ to remove the influence of strand-sampling density and strand length:
\begin{equation}
    \sigma_j = \frac{1}{2} \left( \| \hat{\mathbf{p}}_j - \hat{\mathbf{p}}_{j-1} \| + \| \hat{\mathbf{p}}_{j+1} - \hat{\mathbf{p}}_j \| \right),
\end{equation}

To ensure invariance to global orientation, we represent these normalized deviations in a local orthonormal frame $\mathbf{F}_j$ defined along the coarse backbone.
At each point $\hat{\mathbf{p}}_j$, the frame is constructed using Parallel Transport~\cite{Bishop1975There}.
Together with the scale factor, the style residual at each point is computed as
\begin{equation}
    \mathbf{r}_j = \frac{1}{\sigma_j}\,\mathbf{F}_j^\top (\mathbf{p}_j - \hat{\mathbf{p}}_j).
\end{equation}
Since $\hat{\mathbf{p}}_j$, $\mathbf{F}_j$, and $\sigma_j$ are deterministically defined from the strand, the original strand can be exactly reconstructed from the residuals and coarse backbone, making the parameterization bijective.
This formulation therefore yields a scale- and rotation-invariant representation of local strand texture, allowing residual patterns to be compared and reused across strands as meaningful strand-level ``style words,'' and to be naturally organized as a sequential token stream for autoregressive hairstyle synthesis.

%% file: sec/3_2_hair_model.tex
\section{HairGPT: Unified Autoregressive Hairstyle Synthesis}
\label{sec:hairgpt}

Building upon the dual-decoupled parameterization in Sec.~\ref{sec:parameterization}, we next introduce the generative framework of HairGPT\@. The key idea is to convert the structured hairstyle representation into a discrete, language-like modality that can be modeled autoregressively under multimodal conditions.
To this end, we first construct aligned image-text-hair training data; then we discretize the guide strands, including spatial roots $\mathbf{u}$, coarse geometry $\mathbf{c}$, and style residuals $\mathbf{r}$, as well as the density map, into tokens, organize them into region-aware hierarchical sequences, and model their generation with a decoder-only vision-language model. This section describes how the representation defined in Sec.~\ref{sec:parameterization} is operationalized into a unified multimodal framework for structured hairstyle synthesis.

\subsection{Scalable Multimodal Data Engine}
\label{sec:data_engine}
To address the scarcity of paired 3D hair data, we construct a scalable pipeline that converts heterogeneous sources into aligned triplets $(\mathcal{I}, \mathcal{T}, \mathcal{H})$ of image, text, and hairstyle geometry.

\begin{figure}[t]
  \includegraphics[width=0.95\linewidth]{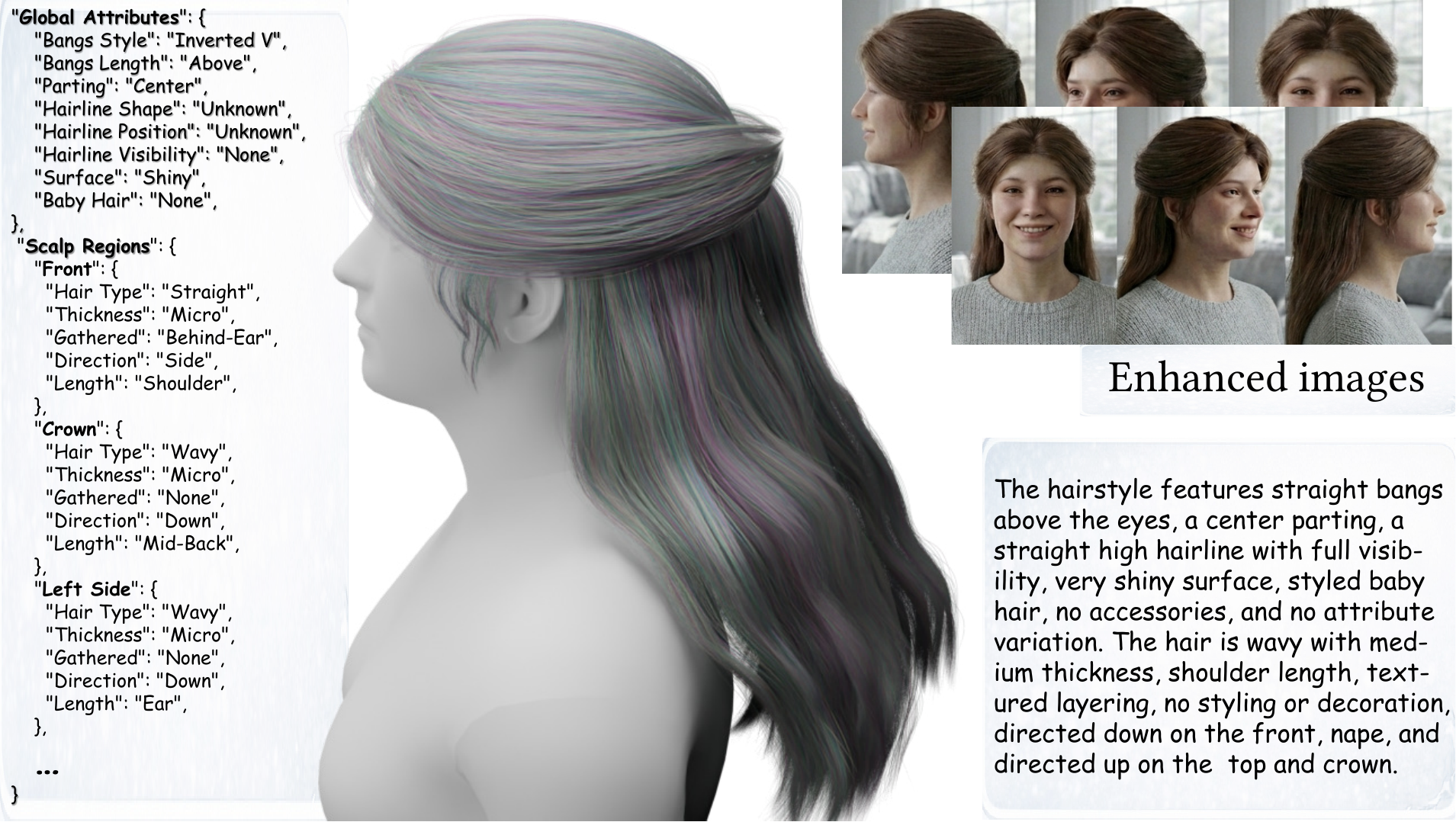}
  \caption{\textbf{Data Example.} For a 3D hair-strand model, we annotate global and local text attributes for distinct scalp regions and provide an overall natural-language hairstyle description. We also utilize a generative model to render diverse photorealistic identities consistent with the underlying hair topology.}
  \Description{Example data pipeline output showing a 3D hair model, region-aware textual annotations, and rendered photorealistic identity variations.}
  \label{fig:data_processing}
\end{figure}

\paragraph{Generative Visual Synthesis ($\mathcal{I}$).}
To bridge the gap between synthetic training data and real-world inference, we leverage a pretrained generative model (e.g., Qwen-Image) to synthesize diverse visual identities, as shown in Fig.~\ref{fig:data_processing}. We prompt the VLM with canonical hair renders and specific attribute instructions, systematically varying skin tones, clothing styles, lighting conditions, and background scenes. This generative augmentation reduces domain bias and improves robustness to real-world visual conditions.

\paragraph{Region-Aware Semantic Annotation ($\mathcal{T}$).}

We observe that large pretrained VLMs already encode rich semantic priors for hairstyle understanding. Beyond recognizing coarse global attributes, they are often capable of describing how a hairstyle is deliberately authored---including how volume, flow, and local styling cues are arranged across different parts of the head to produce a coherent overall design. This makes them a natural source of semantic supervision for our problem, since they already capture much of the human prior needed to verbalize hairstyle structure from images. However, their outputs are typically unstructured and global, and do not explicitly align with localized scalp regions (e.g., distinguishing a fringe from temple hair).

To leverage such pretrained priors while enforcing semantic structure, we further fine-tune  Qwen2.5-VL-7B-Instruct using Low-Rank Adaptation (LoRA) on a curated set of expert-labeled hairstyle images from Hairmony~\cite{Meishvili2024Hairmony}. Concretely, we perform supervised fine-tuning on the Hairmony dataset, where each image is paired with manually annotated structural hair-part attributes, enabling the model to recognize and categorize distinct hair parts from input portraits. We use LoRA adapters with intrinsic rank $r=8$ on all linear modules in the transformer blocks, covering both attention and MLP layers.

The resulting domain-adapted VLM is prompted to generate hierarchical textual annotations that follow the Hairmony taxonomy, producing both a \emph{global hairstyle description} and \emph{region-specific labels} for each of the $M=8$ scalp regions (Sec.~\ref{sec:representation_concept}).
These region-aware captions provide fine-grained semantic supervision that is explicitly aligned with strand geometry, enabling consistent cross-modal learning between text and 3D hair structure, as shown in Fig.~\ref{fig:data_processing}.

\paragraph{Data Aggregation and Geometric Lifting.}
Our dataset combines synthetic priors (DiffLocks~\cite{rosu2025difflocks}, Perm~\cite{he2024perm}), hundreds of artist-authored hairstyles, and reconstructed 3D hair from the Hairmony dataset.
For image-only data, we lift geometry using a diffusion-based reconstruction model and apply VLM-based rejection sampling for quality control.
The final corpus contains approximately \textbf{120k aligned triplets}.

\subsection{Disentangled Geometric Tokenization}
\label{sec:tokenization}

To enable discrete autoregressive generation, we discretize the dual-decoupled representation established in Sec.~\ref{sec:parameterization} through three specialized tokenization processes targeting spatial roots, strand geometry, and the global density map.

\paragraph{Strand Root Quantization ($\mathbf{u} \to \{u, v\}$).}
We discretize the continuous UV coordinates of the root $\mathbf{u} \in \mathcal{S}$ into integer indices. Specifically, we map the 2D scalp manifold $\mathcal{S}$ to a $256 \times 256$ grid. Each guide strand root $\mathbf{u}$ is quantized into a spatial token pair $(u, v)$, where $u, v \in [0, 255]$. These tokens act as the spatial anchors for all subsequent geometric attributes in the generative sequence.

\paragraph{Multi-Head Geometric VQ-VAE ($\mathbf{c}, \mathbf{r} \to T_{\text{coa}}, T_{\text{sty}}$).}
We train two independent encoders for the coarse backbone $\mathbf{c}$ and the style residual $\mathbf{r}$, respectively. 
To enhance codebook expressivity and mitigate codebook collapse, we employ a product quantization strategy with four heads. The latent representations of $\mathbf{c}$ and $\mathbf{r}$ are partitioned into four sub-vectors, each discretized by an independent sub-codebook. This process yields a sequence of four coarse tokens $T_{\text{coa}} = \{c_1, \dots, c_4\}$ and four style tokens $T_{\text{sty}} = \{r_1, \dots, r_4\}$ per strand. 
This compact design uses only eight discrete tokens per strand, which is critical for keeping autoregressive sequence lengths manageable at the scale of hundreds of guide strands; by contrast, more direct parameterizations based on dense curve coefficients would require substantially longer token sequences, making training impractical for hairstyles with 512 guide strands.

The geometric tokenizers are pretrained on high-quality synthetic datasets (DiffLocks and Perm) to learn robust geometric priors, and remain frozen during the LLM training stage to ensure latent stability.

\paragraph{Global Density ($\mathcal{D} \to \mathbf{D}$).}
Beyond individual strands, we encode the density map $\mathcal{D}$ (defined in Sec.~\ref{sec:representation_concept}) using a standard 2D VQ-VAE\@. This autoencoder compresses the dense scalp-space density grid into a $32 \times 32$ latent feature map, which is subsequently quantized into a sequence of $1024$ density tokens $\mathbf{D} = \{d_1, \dots, d_{1024}\}$. These tokens $\mathbf{D}$ serve as a global structural condition, empowering the model to reason about the overall hair-root distribution before synthesizing specific guide strands.

\subsection{Autoregressive Hairstyle VLM}

\subsubsection{Model Architecture and Multimodal Alignment}
We adopt the pretrained Qwen model as our backbone transformer $\Phi$. To bridge the structural hair representation with the linguistic space, we unify all discrete hair components---including UV coordinates (from $\mathbf{u}$), coarse backbones $\mathbf{c}$, style residuals $\mathbf{r}$, and density maps $\mathcal{D}$---into a unified vocabulary $\mathcal{V}$. Each component's local index is mapped to a global token ID $T \in \mathcal{V}$ via component-specific offsets. For any geometric token at step $t$, its input representation $\mathbf{x}_t$ is derived from the learnable embedding table: $\mathbf{x}_t = \mathbf{e}_{\text{geo}}(T_t)$, where $\mathbf{e}_{\text{geo}}$ denotes the embedding lookup function corresponding to the global token ID $T_t$.
The model processes multimodal conditions $\mathcal{C} = \{\mathcal{I}, \mathcal{T}\}$, where $\mathcal{I}$ is a reference image and $\mathcal{T}$ represents linguistic instructions. To incorporate visual guidance, we extract patch-level features from a frozen DINOv3~\cite{simeoni2025dinov3} encoder and map them to the transformer's hidden dimension via a lightweight projector $\mathcal{P}_{\text{img}}$: $\mathbf{E}_{\text{img}} = \mathcal{P}_{\text{img}}(\text{DINOv3}(\mathcal{I}))$.
In contrast, textual prompts $\mathcal{T}$ are processed by retaining Qwen's original word embeddings, which preserves the backbone's pretrained linguistic knowledge. These textual features are dimensionally aligned as:
\begin{equation} \label{eq:text_embed}
    \mathbf{E}_{\text{txt}} = \mathcal{P}_{\text{txt}}(\mathbf{e}_{\text{word}}(\mathcal{T}))
\end{equation}
where $\mathcal{P}_{\text{txt}}$ ensures dimensional consistency. This hybrid embedding strategy allows the transformer $\Phi$ to interpret visual cues through the specialized projector while maintaining linguistic consistency for textual instructions.

\subsubsection{Hierarchical Sequence Construction}

\begin{figure}[t]
  \includegraphics[width=1.0\linewidth]{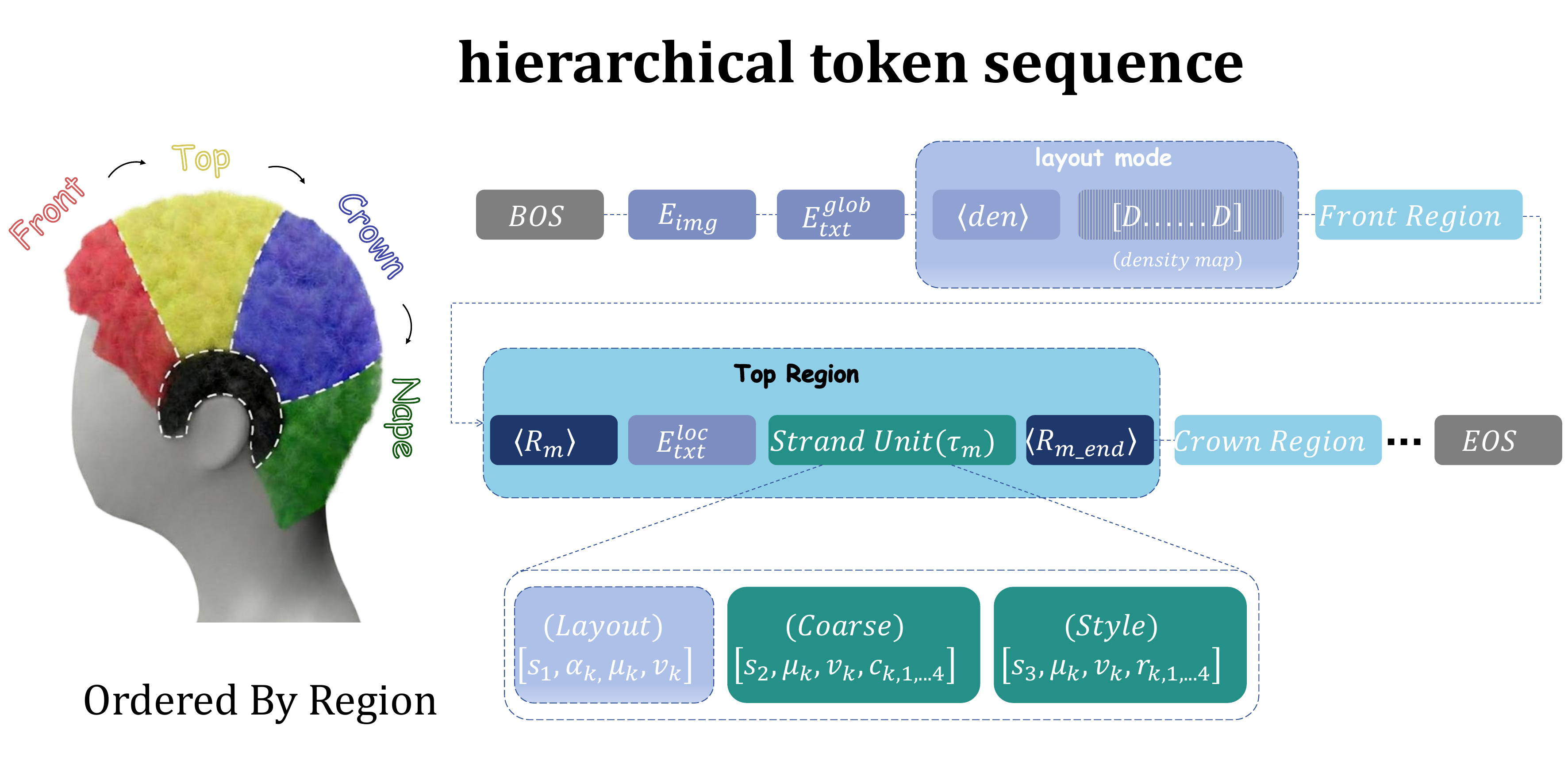}
  \caption{\textbf{Sequence Illustration.} The sequence begins with global text and image embeddings and density map tokens (layout mode). Task-specific strand tokens are generated and constructed region by region, bounded by region markers. Each strand unit is constructed according to the task mode and conditioned on specific separators.}
  \Description{Illustration of the sequence generation process, showing embeddings, tokens, and region markers.}
  \label{fig:sequence}
  
\end{figure}

We linearize a hairstyle into a structured sequence $\mathbf{S}$ that progresses from global layout to strand-level geometric details, enabling autoregressive generation with explicit structural control, as illustrated in Fig.~\ref{fig:sequence}.

\paragraph{Multi-stage Strand Units ($\tau_k$).}
To preserve structural disentanglement, each hair strand $k$ is decomposed into a multi-stage sub-sequence $\tau_k$, organized by task-specific separator tokens $\{s_1, s_2, s_3\}$. 
Conditioned on global context $\mathcal{C}$, the transformer treats spatial anchors $(\alpha_k, u_k, v_k)$ as coordinate queries and predicts the corresponding strand geometry. 
This formulation allows the transformer to be interpreted as a discrete, autoregressive implicit mapping that predicts strand geometry conditioned on spatial queries and global context.

\begin{figure}[t]
  \includegraphics[width=1.0\linewidth]{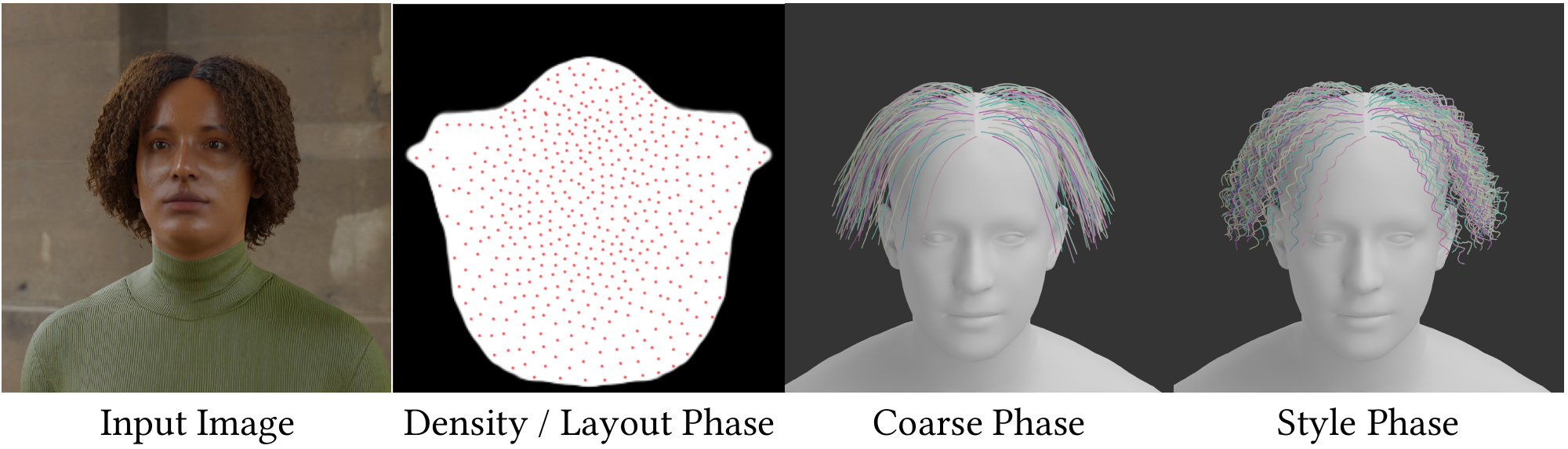}
  \caption{
    \textbf{Phased Autoregressive Generation.} 
    HairGPT progressively generates a hairstyle through multiple phases. 
    The density phase first predicts density tokens, which then condition the following layout phase. Strand-root positions are generated sequentially and visualized as red points. Coarse-strand geometry tokens are then generated, and the style phase finally produces fine-grained residual details. Additional implementation details are provided in the supplementary material.
    }
  \Description{Step-by-step visualization of phased hairstyle generation from density tokens to strand roots, coarse geometry, and fine style residuals.}
  \label{fig:multi-stage-generation}
  
\end{figure}

Specifically, the strand representation is generated through three progressive modes, as illustrated in Fig.~\ref{fig:multi-stage-generation}:
\begin{itemize}
    \item \textbf{Layout Mode:} 
    $\tau_k^{\text{lay}} = [s_1, \alpha_k, u_k, v_k]$, where $\alpha_k$ is a density-aware anchor derived from the density map $\mathcal{D}$. 
    This stage instantiates strand root locations by mapping the global density prior to precise coordinates on the scalp manifold.
    
    \item \textbf{Coarse Mode:} 
    $\tau_k^{\text{coa}} = [s_2, u_k, v_k, c_{k,1}, \dots, c_{k,4}]$, which predicts the low-frequency backbone geometry $\mathbf{c}_k$ conditioned on the spatial query $(u_k, v_k)$.
    
    \item \textbf{Style Mode:} 
    $\tau_k^{\text{sty}} = [s_3, u_k, v_k, r_{k,1}, \dots, r_{k,4}]$, which generates high-frequency style residuals $\mathbf{r}_k$ with spatial tokens.
\end{itemize}

By reusing the same spatial coordinates $(u_k, v_k)$ across all stages, the layout stage strictly determines strand placement, while the coarse and style stages are constrained to only generate geometry conditioned on this fixed spatial anchor.

\paragraph{Region-Aware Assembly.}
Strands are grouped by scalp regions $\mathcal{R}_m$ and assembled into a mode-conditioned sequence:
\begin{equation}
\begin{split}
    \mathbf{S}^{(\cdot)} = \Big[ 
    &\text{BOS},\ \mathbf{E}_{\text{img}},\ \mathbf{E}_{\text{txt}}^{\text{glob}},\ 
    \langle \text{den} \rangle,\ \mathbf{D}, \\
    &\dots,\ \langle \mathcal{R}_m \rangle,\ 
    \mathbf{E}_{\text{txt}, m}^{\text{loc}},\ 
    \tau^{(\cdot)}_{m,1\dots N_m},\ 
    \langle \mathcal{R}_m\text{\_end} \rangle, \\
    &\dots,\ \text{EOS} 
    \Big], \tau^{(\cdot)}_k \in \{\tau^{\text{lay}}_k, \tau^{\text{coa}}_k, \tau^{\text{sty}}_k\}
\end{split}
\end{equation}

where $\mathbf{S}^{(\cdot)}$ denotes one of the layout, coarse, or style sequences.
$\text{BOS}$ and $\text{EOS}$ denote sequence boundary tokens.
The global and region-specific text embeddings are aligned via Eq.~\ref{eq:text_embed}.
The markers $\langle \mathcal{R}_m \rangle$ and $\langle \mathcal{R}_m\text{\_end} \rangle$ delimit semantic regions.
The fixed region order is \texttt{Front}, \texttt{Top}, \texttt{Crown}, \texttt{Nape}, followed by the left and right sides and temples, as shown in Fig.~\ref{fig:sequence}.

\begin{figure*}[t]
	\centering
	\includegraphics[width=1.0\linewidth]{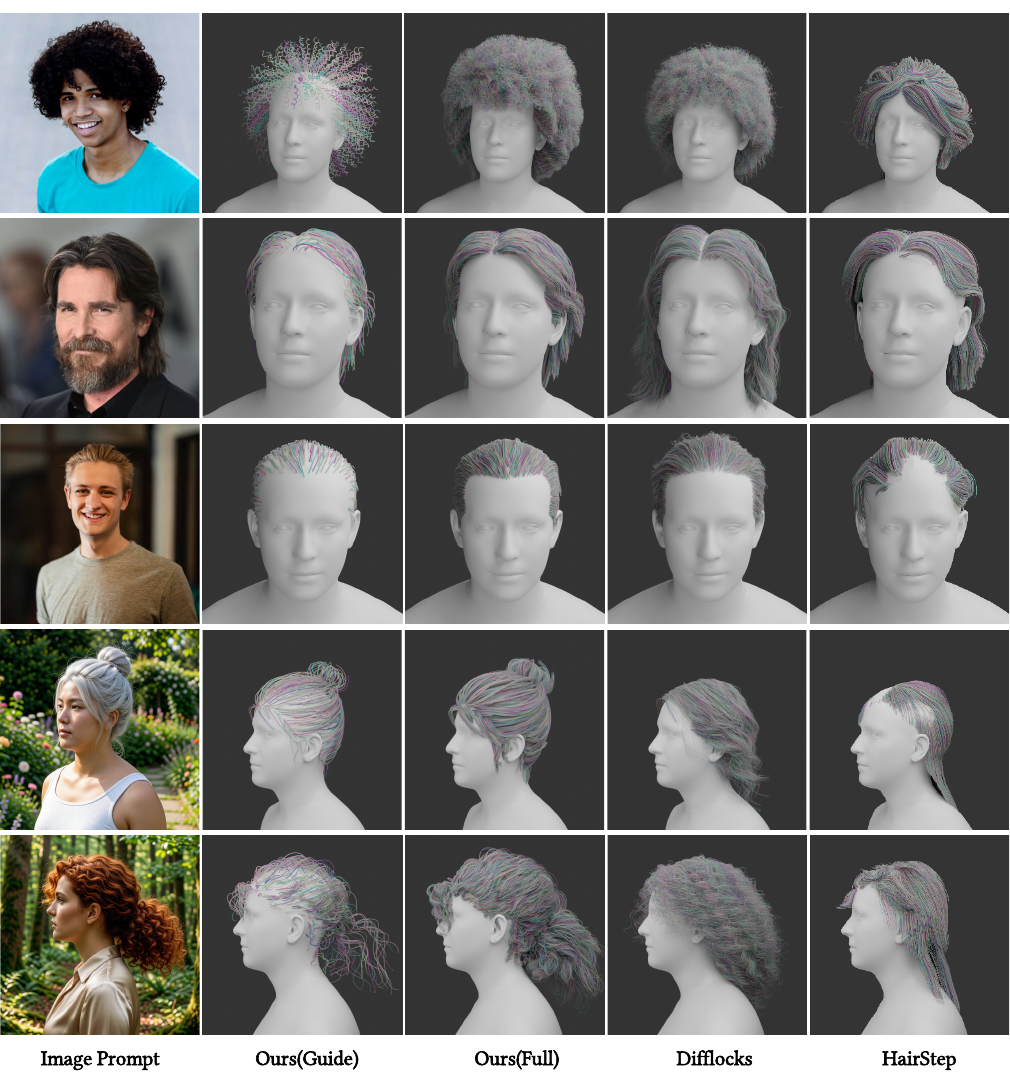}
  \caption{\textbf{Image-guided hairstyle synthesis comparison.} HairGPT effectively generates tightly coiled hairstyles and complex hair topology conditioned on the input image, especially for buns and ponytails. We visualize both the raw guide strands directly output by our model and the dense strands produced via a simple interpolation algorithm; note that this upsampling process is employed solely for visualization and is not the primary focus of this work.}
  \Description{Comparison of image-guided hairstyle synthesis results, showing input images, baseline outputs, HairGPT guide strands, and interpolated dense strands.}
	\label{fig:img2hair_comp}
\end{figure*}

\begin{figure*}[t]
	\centering
	\includegraphics[width=1.0\linewidth]{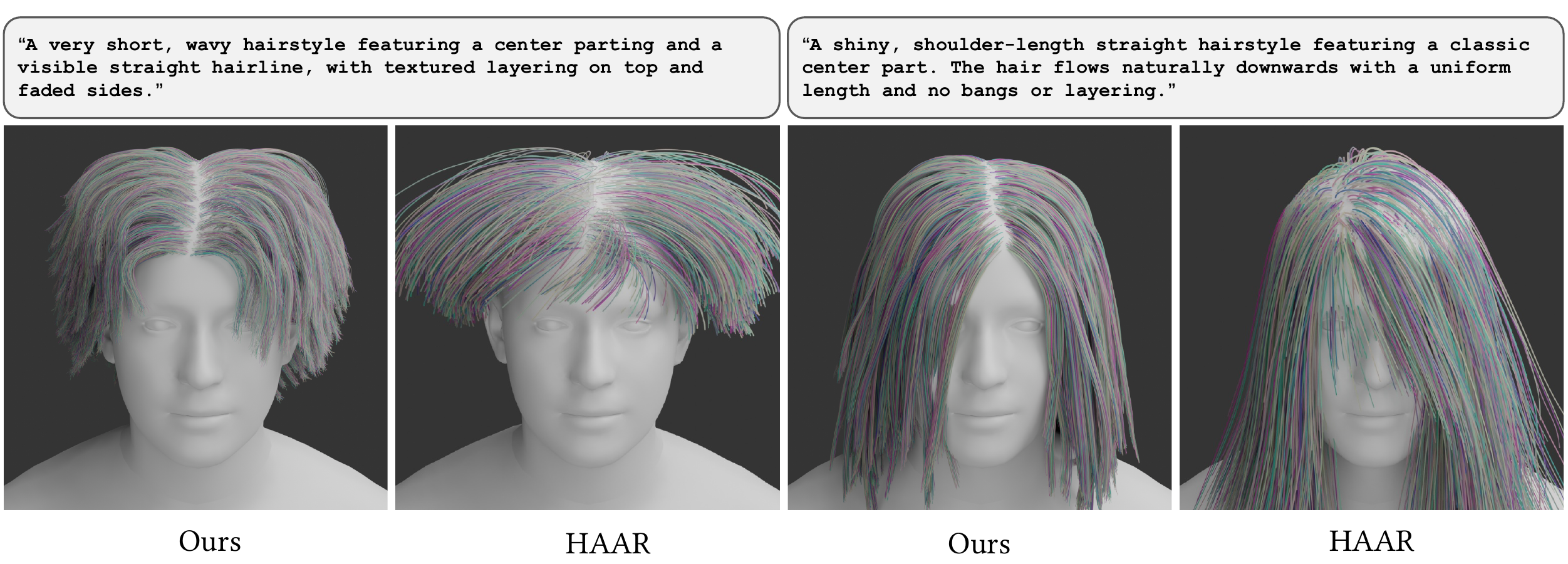}
	\caption{\textbf{Text-guided hairstyle synthesis comparison.} Our HairGPT produces 3D hairstyles that adhere to fine-grained semantic instructions.}
  \Description{Comparison of text-guided hairstyle synthesis results demonstrating HairGPT outputs that follow detailed text prompts.}
	\label{fig:txt2hair_comp}
\end{figure*}

\subsubsection{Multi-Stage Training Strategy}

\paragraph{Mode-Specific Sequence Sampling.} 
During training, each sequence $\mathbf{S}^{(\cdot)}$ is constructed from a \emph{single generation mode}---layout, coarse, or style---rather than concatenating all strand units. 
This ensures the model learns each stage independently while conditioning on the relevant global and regional context. 

\paragraph{Training Loss.} 
The model is trained via next-token prediction with a redundancy mask $\mathcal{M}$ to exclude repeated spatial tokens $(u_k, v_k)$ from the loss in the coarse and style stages. Let $T_t$ denote the token at step $t$ and $\omega(T_t)$ its category-aware weight:
\begin{equation}
\begin{split}
    \mathcal{L} = - \frac{1}{N_{\text{valid}}} \sum_{t} \Big[ & \omega(T_t) \cdot \mathbb{1}(T_t \notin \mathcal{M}) \\
    & \cdot \log P(T_t | T_{<t}, \mathcal{C}) \Big].
\end{split}
\end{equation}
Repeated spatial tokens serve purely as conditioning anchors and are excluded from gradient computation.
Multimodal dropout is applied concurrently to further improve robustness to variations in visual and textual prompts.

%% file: sec/4_results.tex
\section{Results}

We evaluate our model by comparing it with hair generation baselines, key components, and training strategies. Implementation and inference details are provided in the supplementary material.

\subsection{Comparisons}
We compare HairGPT against state-of-the-art hair synthesis baselines, including diffusion-based methods (\textit{DiffLocks}~\cite{rosu2025difflocks}) and reconstruction methods (\textit{HairStep}~\cite{zheng2023hairstep}) for image-guided generation. We also compare our approach with \textit{HAAR}~\cite{sklyarova2023haar} in terms of text-guided synthesis capability.

\paragraph{Image-guided Generation.} We provide only a single reference image for both HairGPT and the baseline methods. While HairStep overlooks fine-grained local hair details and DiffLocks struggles to synthesize hairstyles with complex topological structures, such as ponytails, as illustrated in Fig.~\ref{fig:img2hair_comp}, our framework successfully synthesizes a diverse spectrum of hairstyles---ranging from high-frequency afro hairstyles to complex topological structures, such as buns. Leveraging our dual-decoupled representation, our model maintains superior structural consistency and detail preservation. 

\paragraph{Text-guided Generation.} We provide only text prompts for our HairGPT and HAAR\@.
As illustrated in Fig.~\ref{fig:txt2hair_comp}, our method produces 3D hairstyles that adhere to fine-grained semantic instructions---such as precise parting locations and specific regional styles (e.g., diagonal bangs)---and exhibits superior structural coherence compared to HAAR, benefiting from the language priors inherited from the backbone model.

\begin{table}[t]
\centering
\small
\caption{
CLIP score analysis based on the similarity between the input image condition and rendered 3D hairstyles.}
\label{tab:clip_score_img}
\resizebox{0.7\linewidth}{!}{
\begin{tabular}{lccc}
\toprule
Method & DiffLocks & HairStep & Ours \\
\midrule
CLIP Score $\uparrow$ & 0.577 & 0.537 & \textbf{0.592} \\
\bottomrule
\end{tabular}
}
\end{table}

\begin{table}[t]
\centering
\small
\caption{Quantitative comparison on the DiffLocks evaluation set.}
\label{tab:hair_metrics}
\resizebox{1.0\linewidth}{!}{

\begin{tabular}{lccccc}
\toprule
Method
& CD $\downarrow$ $(\times 10^{-3})$
& DCT-FID $\downarrow$ $(\times 10^{-3})$
& Precision $\uparrow$
& Recall $\uparrow$
& F-score $\uparrow$ \\
\midrule
Ours
& \textbf{6.31}
& \textbf{8.80}
& 0.824
& \textbf{0.833}
& \textbf{0.828} \\
DiffLocks
& 7.83
& 9.97
& \textbf{0.839}
& 0.809
& 0.824 \\
\bottomrule
\end{tabular}
}
\end{table}

\paragraph{Quantitative analysis.}

We also evaluate the generation quality of our model quantitatively. 
First, we report CLIP scores in Tab.~\ref{tab:clip_score_img} and compare against DiffLocks and HairStep.
Without ground-truth (GT) geometry, we compute the CLIP scores based on the feature similarity between the input condition and the rendered 3D hairstyles, using the same lighting, camera views, and shaders for all methods. 
We note that the CLIP score may overemphasize local texture cues while overlooking the global topology of hair geometry, which can slightly underestimate the performance of our method. 
We further evaluate geometric reconstruction quality on the DiffLocks evaluation set. In Tab.~\ref{tab:hair_metrics}, we report point-cloud Chamfer distance and Precision/Recall/F-score metrics between the generated hairstyles and GT strands. 
We also compute the mean DCT-FID by representing each strand with DCT coefficients and measuring the Fr\'echet distance between the generated and GT strand distributions of all hairstyles.

\subsection{Ablation Study}

\paragraph{Multi-head Strand Tokenizer.} 
We evaluate the effectiveness of the tokenizer design with the same parameter budget (0.26M). As shown in Tab.~\ref{tab:tok_ablation}, a baseline \textit{Single-head} exhibits limited capability and lower codebook usage, while the results of our multi-head design (Rows 2--3) indicate that product quantization improves the capacity of the codebook.

\begin{figure}
  \includegraphics[width=1.0\linewidth]{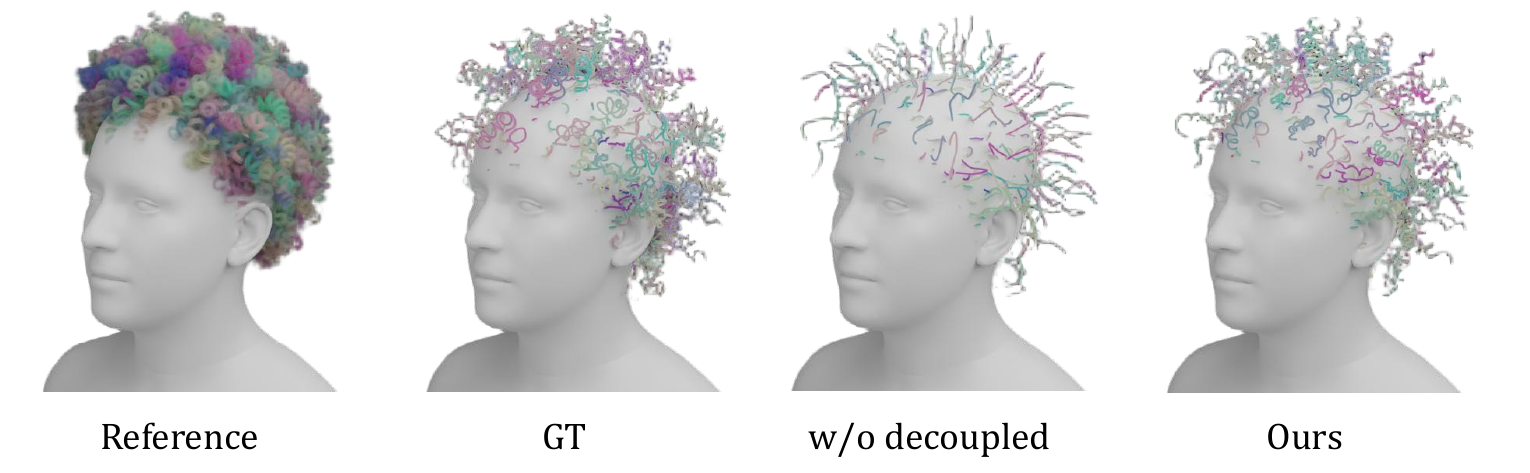}
  \caption{Ablation of strand-level Coarse-Style Decoupling. The decoupling effectively enables the tokenizer to encode high-frequency details.} 
  \Description{Ablation figure comparing strand reconstruction results with and without coarse-style decoupling, highlighting high-frequency detail preservation.}
  \label{fig:ablation_tokenizer}
  
\end{figure}

\paragraph{Coarse-Style Decoupling.} We further evaluate the necessity of the decoupled representation (Sec.~\ref{sec:representation_concept}). As illustrated in Fig.~\ref{fig:ablation_tokenizer}, a tokenizer trained on raw strands directly (\textit{w/o c/r}) struggles to recover high-frequency curling patterns, and fails to capture the characteristic tight coils of the reference afro-texture. Quantitatively, Tab.~\ref{tab:tok_ablation} (Row~2) shows that without coarse-style decomposition, the model suffers from significantly higher directional error and positional drift. In contrast, our full model can disentangle high-frequency residuals ($\mathbf{r}$) from the low-frequency topological backbone ($\mathbf{c}$).

\begin{figure}
  \includegraphics[width=1.0\linewidth]{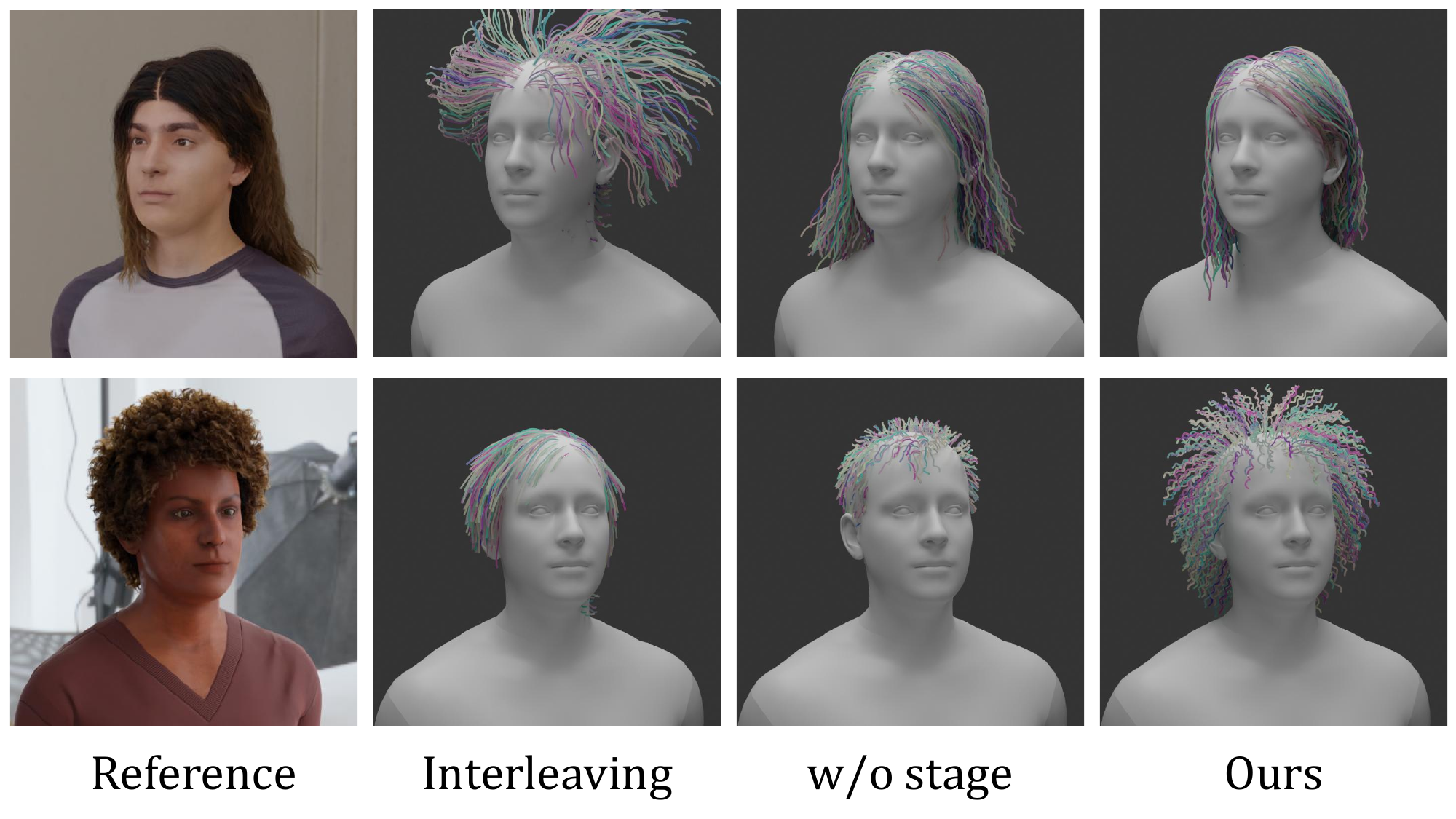}
  \caption{Ablations on sequence construction and multi-stage training. }
  \Description{Ablation figure showing hairstyle generation differences under alternative sequence construction and training strategies.}
  \label{fig:ablation_sequence}
\end{figure}

\paragraph{Impact of Sequence Construction.} 
We evaluate the role of task-specific separators by evaluating an \textit{Interleaving} baseline, where the root ($u$), coarse ($c$), and style ($r$) tokens of all strands are directly concatenated into a continuous stream. As shown in Fig.~\ref{fig:ablation_sequence}, this configuration leads to complete geometric collapse. Without the anchoring effect of separators to clarify the task, the model suffers from severe autoregressive drift; the error accumulates through the subsequent geometry tokens, resulting in strands that are scattered chaotically and fail to form a coherent hairstyle.

\paragraph{Multi-stage Training.} 
Furthermore, we evaluate the importance of our $Layout/Coarse/Style$ staged training strategy. In the \textit{w/o stage} variant, the model is trained on a single, flattened sequence containing all geometric layers at once. While this baseline can capture basic silhouettes for low-frequency straight hair, it struggles significantly with high-frequency afro-textures (bottom row), due to the long-range dependency bottleneck in extremely long sequences ($N=512$ strands, about 10k tokens). Without staged supervision, the high-frequency style details at the end of the sequence suffer from information decay and attention attenuation relative to the global conditions. This results in the sparse, fragmented strands observed in Fig.~\ref{fig:ablation_sequence}.

\begin{table}[t]
\small
\centering
\caption{Ablations on tokenizer and strand-level Coarse-Style Decoupling.}
\label{tab:tok_ablation}
\resizebox{1.0\linewidth}{!}{
\begin{tabular}{lcccccc}
\toprule
Method & Pos $\downarrow$ & Dir $\downarrow$ & Curv $\downarrow$ & Usage (\%) $\uparrow$ & Heads $\uparrow$ & \#Params \\
\midrule
Single-head                 & $4.5 \times 10^{-3}$ & $1.2 \times 10^{-2}$ & $6.3 \times 10^{-4}$ & 63.9 & 1 & 0.26M \\
W/o $\mathbf{c}/\mathbf{r}$ & $2.5 \times 10^{-3}$ & $2.0 \times 10^{-2}$ & $6.5 \times 10^{-4}$ & \textbf{99.1} & 4 & 0.26M \\
Ours-full                   & $\mathbf{1.2 \times 10^{-3}}$ & $\mathbf{3.4 \times 10^{-3}}$ & $\mathbf{6.3 \times 10^{-4}}$ & {98.9} & 4 & 0.26M \\
\bottomrule
\end{tabular}
}
\end{table}

\subsection{Applications}

\begin{figure*}[t]
	\centering
	\includegraphics[width=1\linewidth]{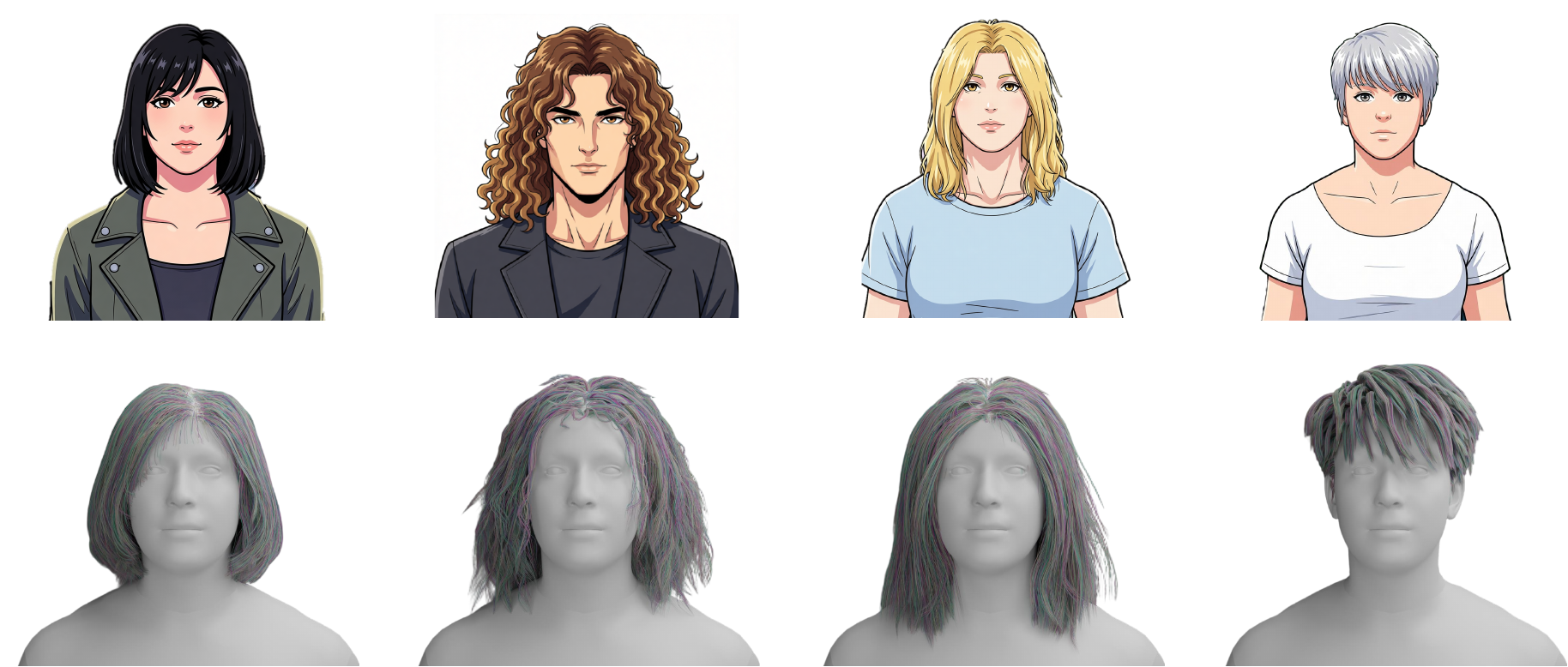}
	\caption{\textbf{Cross-domain adaptation to stylized characters.} Our framework adapts to 2D cartoon inputs via fine-tuning. It generates plausible 3D strand arrangements that faithfully respect the volume and flow of the original anime portraits.}
  \Description{Examples of adapting HairGPT to stylized anime character inputs and producing corresponding 3D hairstyle strand structures.}
	\label{fig:cart2hair}
\end{figure*}

\paragraph{Cross-Domain Adaptation.} 
Although HairGPT is pretrained on realistic data, the structural priors it acquires facilitate efficient domain adaptation via fine-tuning. Fig.~\ref{fig:cart2hair} illustrates the results on 2D anime characters: the model infers plausible hair topology from flat-shaded images, preserving the distinct aesthetic characteristics of the input styles. This capability highlights the potential of our representation to serve as a unified geometric backbone that can be specialized for diverse artistic domains with limited additional data.

\begin{figure*}[t]
	\centering
	\includegraphics[width=1\linewidth]{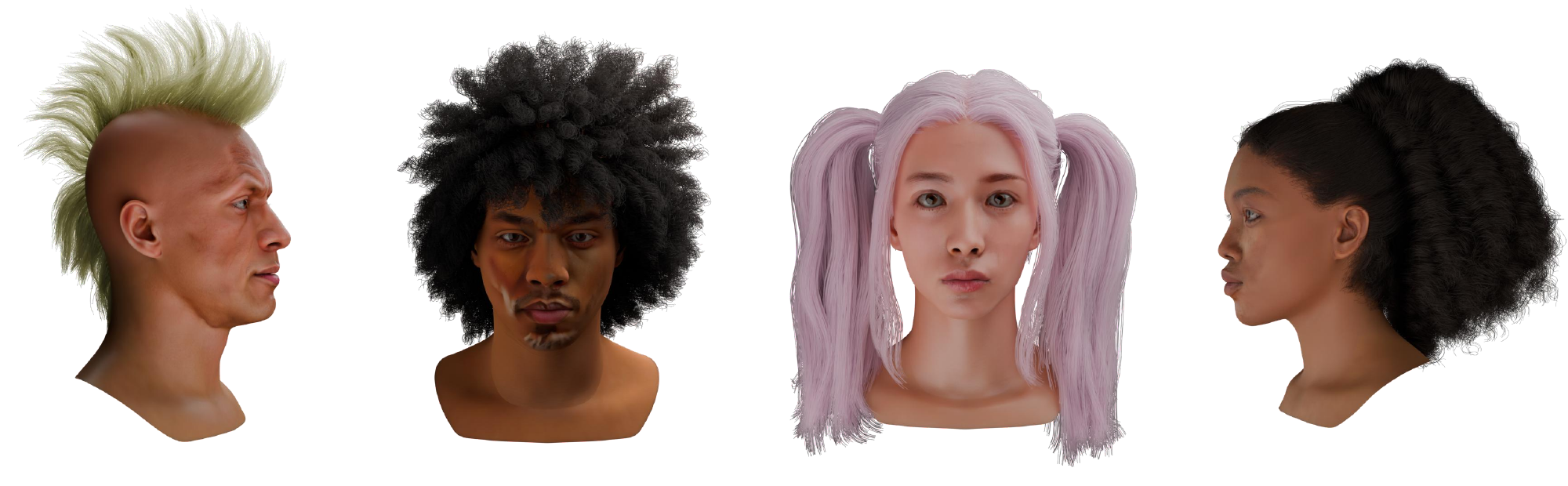}
	\caption{\textbf{Realistic Avatar Creation.} Our model can work in conjunction with the 3D face synthesis model~\cite{dreamface} to produce photorealistic avatars with unified visual aesthetics.}
  \Description{Examples of photorealistic avatar creation combining generated 3D hair with 3D face synthesis results.}
	\label{fig:dreamface}
\end{figure*}

\paragraph{Realistic Avatar Creation.} 
Conditioned on compatible semantic attributes (e.g., specific eras or cultural styles), HairGPT can work in conjunction with the 3D face synthesis model~\cite{dreamface} to produce photorealistic avatars with unified visual aesthetics (Fig.~\ref{fig:dreamface}). Crucially, this approach maintains geometric disentanglement, providing high-quality, separable hair and face assets that are ready for independent animation and editing.

\begin{figure*}[t]
	\centering
	\includegraphics[width=1\linewidth]{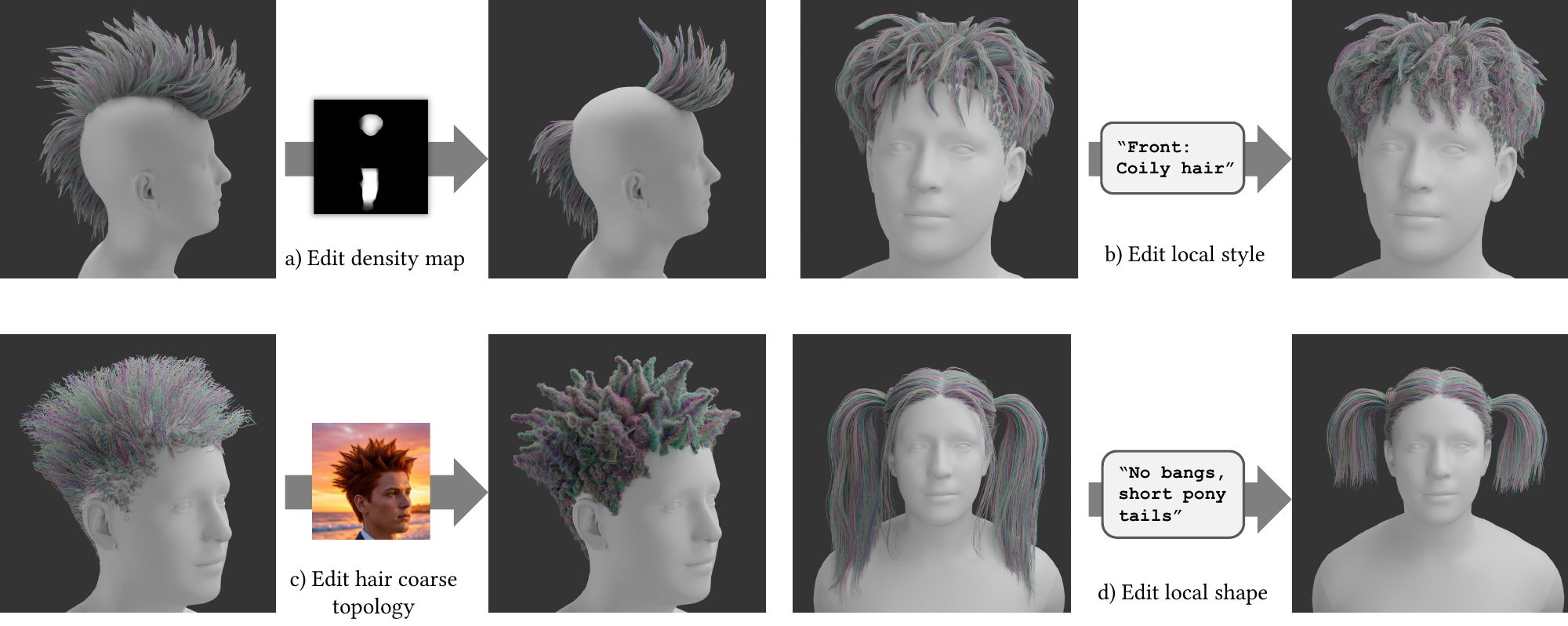}
	\caption{\textbf{Editing.} Our dual-decoupled representation and vision-language model naturally facilitate diverse editing applications with either image or text prompts.}
  \Description{Editing examples showing changes to hair density, coarse shape, and local style using image or text prompts.}
	\label{fig:editing}
\end{figure*}

\paragraph{Flexible Multimodal Editing.} 
Our dual-decoupled representation and vision-language model naturally facilitate diverse editing applications. As shown in Fig.~\ref{fig:editing}, users can explicitly control hair volume and distribution by modifying the density map (a). Thanks to the disentanglement of topology and texture, the coarse geometry can be guided by a reference image to transfer global styles (c), while local details---such as specific curl patterns (b) or regional shapes like bangs (d)---can be precisely altered via regional text prompts, without disrupting the overall hair structure.

\begin{figure}[t]
	\centering
	\includegraphics[width=1\linewidth]{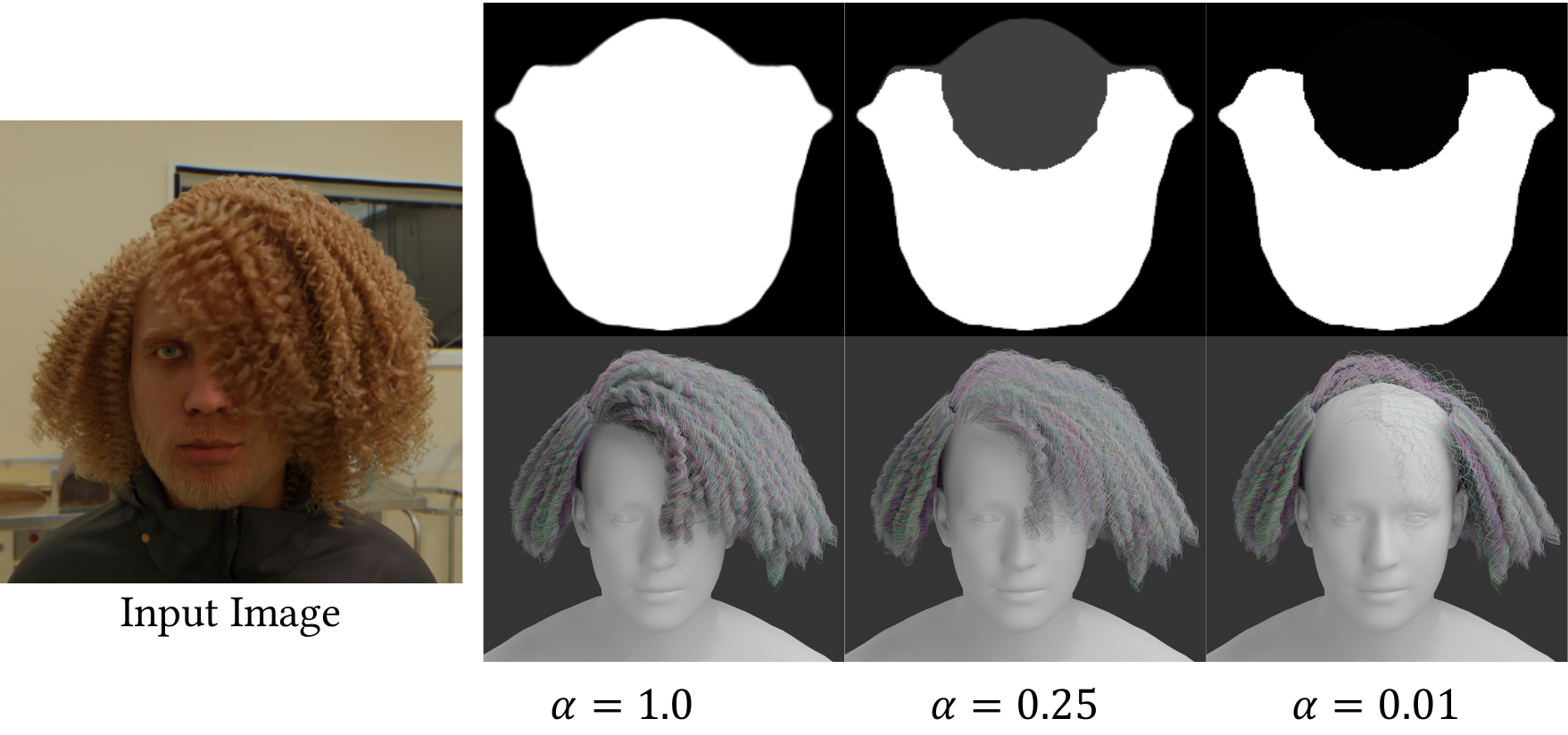}
	\caption{By tuning the scaling factor $\alpha$, we can continuously control the hair density in the top region.}
  \Description{Visualization of continuous hair-density control in the top scalp region as the scaling factor changes.}
	\label{fig:density_scale}
\end{figure}

\subsection{Limitations and Discussion}
\paragraph{Limitations.}
Despite the strong controllability and semantic expressiveness enabled by our strand-centric formulation, the current framework still has several limitations. First, our data engine relies in part on a fine-tuned vision-language model for scalable semantic annotation and augmented supervision. While this greatly improves data construction efficiency, the resulting labels still do not fully match expert human annotation, especially for subtle regional attributes, ambiguous boundaries, and rare hairstyle patterns. Incorporating more human-curated annotations and stronger human-in-the-loop verification would further improve the pipeline's reliability.

Second, strand tokenization remains inherently challenging. Our compact design represents each strand using only eight discrete tokens, which is critical for making autoregressive modeling feasible at the scale of hundreds of guide strands. However, this compactness inevitably sacrifices some local geometric fidelity. Although reconstruction quality is sufficient in our setting, strands with extremely high curvature or irregular local textures may be smoothed. Richer local parameterizations, adaptive tokenization, or hybrid detail representations may help alleviate this limitation.

Third, our framework focuses on generating structurally meaningful guide strands rather than dense hair directly. Dense hairstyles are currently obtained through a simple interpolation procedure, and the final visual quality may therefore be affected by the interpolation algorithm itself. A natural future direction is to combine our autoregressive guide-strand representation with a dedicated downstream refinement model for dense strand synthesis. In addition, autoregressive inference (30--60 seconds) remains slower than diffusion-based alternatives because strand-level generation requires long token sequences. Improving efficiency through more compact representations or more parallel generation strategies remains an important future direction.

\paragraph{Discussion.}
We believe the significance of HairGPT lies not only in introducing a new model for hairstyle synthesis, but more importantly in advancing a new generative paradigm for hair. Rather than treating hairstyle as a single entangled field to be synthesized holistically, our framework represents it as an explicit, structured, and semantically grounded composition of strands and formulates generation as an autoregressive construction over layout, coarse structure, and fine style. This makes the generative process more transparent, more aligned with the structured logic of hairstyle authoring, and better suited for translating high-level intent into coherent 3D geometry.

More importantly, this strand-centric formulation opens a broad space of new problems beyond static generation. Because strands remain explicit throughout the pipeline, the same representation naturally supports semantic editing, topology transfer, hairstyle completion, sparse-to-dense grooming, personalized avatar creation, and simulation-aware generation. We therefore view HairGPT not as an endpoint, but as an initial foundation for a broader family of structured hair generation systems.

Looking forward, we believe this formulation also provides a promising foundation for more agentic hairstyle generation. Rather than producing all strands in a single pass, future systems could iteratively plan, edit, and verify hairstyle structure across semantic scalp regions and generative stages, using high-level goals to guide global layout before refining local strand style. Such an agentic formulation would move beyond one-shot generation toward interactive and self-refining 3D hair authoring. We hope this work can motivate future research on compositional, artist-aligned, cognitively meaningful, and ultimately agentic generative models for complex 3D content.

%% file: sec/5_conclusion.tex
\section{Conclusion}

We proposed HairGPT, a strand-centric framework that models realistic 3D hairstyles through dual-decoupled autoregressive generation.
At its core is the dual-decoupled hairstyle representation, accompanied by a carefully designed geometric tokenizer that enables effective strand discretization. 
HairGPT transforms hair generation from opaque texture synthesis into a transparent, structured, and semantically meaningful process. 
Experiments demonstrate that this framework supports robust semantic conditioning and compositional editing, enabling the high-fidelity generation of rare, complex hairstyles and effective downstream adaptation to stylized domains.

%% file: sec/supplementary.tex
\section{Architecture Details of Strand Tokenizer}
\label{sec:strand_tokenizer_arch}

The strand tokenizer is implemented as a Vector Quantized Variational Autoencoder (VQ-VAE) tailored for sequential 1D geometric data. The detailed architecture is described below.

\subsection{Network Structure}
Given that hair strands are represented as ordered sequences of 3D coordinates, we leverage 1D Convolutional Neural Networks (1D-CNNs) as the backbone.

\begin{itemize}
    \item \textbf{Encoder:} The encoder transforms the input strand geometry $\mathbf{x} \in \mathbb{R}^{N \times 3}$ into a latent representation. It consists of stacked 1D convolutional blocks. To ensure training stability, we apply Weight Normalization across layers, followed by non-linear activations (e.g., LeakyReLU). The network progressively reduces the temporal resolution, resulting in a latent tensor $\mathbf{z}_e \in \mathbb{R}^{4 \times 8}$.
    
    \item \textbf{Decoder:} The decoder mirrors the encoder's architecture. It utilizes transposed 1D convolutions to upsample the quantized latent codes back to the original spatial resolution $N$, reconstructing the strand geometry $\hat{\mathbf{x}}$.
\end{itemize}

\subsection{Multi-Head Vector Quantization}
To balance codebook usage and reconstruction fidelity, we employ a Multi-Head Vector Quantizer (Product Quantization) mechanism.

\begin{itemize}
    \item \textbf{Subspace Decomposition:} The latent dimension $D$ is naturally set to 8, and the tokenizer contains 4 heads.
    
    \item \textbf{Independent Codebooks:} We maintain $4$ separate codebooks $\{\mathcal{C}_1, \dots, \mathcal{C}_4\}$. Each codebook $\mathcal{C}_m$ contains $K$ entries. Quantization for the $m$-th head is performed by finding the nearest neighbor in $\mathcal{C}_m$. We use cosine similarity as the distance metric, as it provides higher codebook utilization than standard Euclidean distance in our experiments. We use 8192 entries for the coarse-shape VQ-VAE and 2048 entries for the style-residual VQ-VAE.
    
    \item \textbf{Training Stability:} The model is trained using a combination of reconstruction loss and commitment loss. To further improve stability:
    \begin{enumerate}
        \item \textbf{EMA Updates:} For codebook updates, we use Exponential Moving Average (EMA), which smooths the learning dynamics.
        \item \textbf{Noise Injection:} We inject scaled random noise into the latent codes during training: $\mathbf{z}_{noisy} = \mathbf{z}_e + \epsilon$, where $\epsilon \propto \text{std}(\mathbf{z}_e)$. This noise injection encourages the encoder to generate robust representations and prevents the quantizer from relying on precise floating-point artifacts.
    \end{enumerate}
\end{itemize}

\subsection{Efficient Training via Cluster-based Sampling}
\label{sec:sampling_strategy}

To facilitate efficient training and handle the intractable scale of the original synthetic datasets (typically containing $\sim 10^5$ strands per model), we adopt a stratified sampling approach based on our clustering results. 

Following the frequency-aware clustering described in the main paper, which partitions the hair volume into $N_{\text{guide}} = 512$ clusters, we do not train on the full dense geometry. Instead, for each cluster, we randomly sample 10 individual strands as training representatives. This sampling strategy allows the model to learn the local geometric variance within each cluster while reducing the total number of strands per hairstyle to a manageable size ($5,120$ strands). This approach ensures a balanced representation of all semantic regions of the scalp and maintains the structural richness of the original hairstyle with significantly lower computational overhead.

\section{Data Details}

\subsection{Text Annotation Example.}
To illustrate the granularity of our data engine, we provide a representative example of the textual annotations associated with a 3D hairstyle. Our annotations consist of two components: structured hierarchical attributes and a synthesized natural language description.

\paragraph{Structured Attributes.} 
The metadata is organized into global hairstyle characteristics and region-specific geometric properties for $M=8$ scalp regions.

\begin{itemize}
    \item \textbf{Global Attributes:} 
    \begin{itemize}
        \item \textit{Parting:} Center; \textit{Bangs:} None; \textit{Surface:} Shiny.
        \item \textit{Hairline:} Straight shape, Medium position, Full visibility.
        \item \textit{Others:} No accessories, No baby hair, No attribute variation.
    \end{itemize}
    
    \item \textbf{Scalp Region Details:}
    \begin{description}
        \item[Front, Sides, Temples:] Straight hair type, medium thickness, directed \textbf{downward}, chin length. Layering is textured (except at temples).
        \item[Top, Crown:] Straight hair type, medium thickness, directed \textbf{upward}, chin length, textured layering.
        \item[Nape:] Straight hair type, medium thickness, directed \textbf{downward}, \textbf{short} length, textured layering.
    \end{description}
\end{itemize}

\paragraph{Synthesized Natural Language Description.} 
The domain-adapted VLM converts the structured attributes above into a coherent, semantically rich paragraph used for cross-modal training:

\begin{quotation}
\noindent \textit{``The hairstyle features no bangs, no accessories, a center parting, a straight medium-positioned hairline with full visibility, a shiny surface, no baby hair, and consistent attributes throughout. The hair is straight with medium thickness, directed downward in the front, right side, left side, nape, and right and left temples, and upward in the top and crown. Lengths reach the chin in most areas (front, top, crown, right side, right temple, left temple, left side) and are short at the nape. Layering is textured in the front, top, crown, right side, and left side, while the right and left temples have no layering.''}
\end{quotation}

This hierarchical annotation ensures that HairGPT learns to associate specific linguistic tokens with localized geometric structures, such as the upward growth flow at the crown versus the downward flow at the nape.

\subsection{Multimodal Annotation and Augmentation Pipeline}

To construct the high-quality aligned triplets $(\mathcal{I}, \mathcal{T}, \mathcal{H})$ required for training \textit{HairGPT}, we developed a multi-stage data engine. This pipeline performs hierarchical semantic labeling, natural language normalization, and generative visual augmentation to bridge the gap between synthetic geometry and real-world diversity.

\paragraph{Phase 1: Hierarchical Semantic Labeling.} 
We first extract fine-grained attributes from 3D hair models using a domain-adapted Qwen2.5-VL-7B-Instruct model. As shown in Listing~\ref{lst:vision_prompt}, we employ a strict \textbf{Hairstyle Taxonomy Prompt} that enforces a structured JSON output. This taxonomy covers $10$ global attributes (e.g., parting type, hairline shape) and $8$ localized geometric parameters for each of the $M=8$ scalp regions. By restricting the VLM to a predefined candidate set, we ensure taxonomic consistency across the dataset.

\paragraph{Phase 2: Natural Language Normalization.} 
To convert structured JSON metadata into natural-language descriptions suitable for transformer training, we use a \textbf{Normalization System Prompt}. The model is instructed to synthesize a coherent narrative starting with \textit{``The hairstyle features...''}, prioritizing global style before detailing regional specifics. This process serializes spatial attributes into a semantic modality that is rich in contextual information.

\paragraph{Phase 3: Contextual and Stylized Augmentation.} 
To improve robustness, we implement a generative augmentation pipeline. Using the extracted attributes, we systematically vary $13$ hair colors, $12$ subject identities, $11$ clothing types, and $12$ backgrounds to synthesize augmented images.
\begin{itemize}
    \item \textbf{Photorealistic Branch:} Focused on creating diverse real-world contexts (e.g., \textit{cyberpunk street}, \textit{modern office}) to prevent overfitting to synthetic renders.
    \item \textbf{Stylized Branch (Anime):} Aimed at cross-domain generalization. We utilize specific prompts for \textit{cel-shading} and \textit{2D flat colors} to create anime counterparts while preserving the exact 3D hair topology (see Listing~\ref{lst:aug_prompt}).
\end{itemize}

\begin{lstlisting}[language=json, caption={Fragment of the Hairstyle Taxonomy Prompt.}, label=lst:vision_prompt, basicstyle=\tiny\ttfamily, frame=single, breaklines=true]
Analyze the hairstyle... Fill each field using only values from:
Global Attributes:
- Bangs Style: [None, Straight, V-shape, Inverted U, ...]
- Parting: [Center, Right, Left, Zig-Zag, ...]
Scalp Regions (Front, Top, Crown, Nape, ...):
- Hair Type: [Coily, Curly, Wavy, Straight]
- Direction: [Down, Side, Up, Out]
- Length: [Very Short, Ear, Chin, Shoulder, ...]
Return ONLY JSON: 
{ "Global Attributes": {...}, "Scalp Regions": {"Front": {...}, ...} }
\end{lstlisting}

\begin{lstlisting}[caption={Image Augmentation Prompt Templates.}, label=lst:aug_prompt, basicstyle=\tiny\ttfamily, frame=single, breaklines=true]
# Realistic Photo Template:
"A realistic photo of {subject} {clothing}, {background}. The person has {hair_color}. Keep the exact same hairstyle and structure as the input image. High quality, photorealistic."

# Anime Style Template:
"High quality anime style illustration of {subject} {clothing}, {background}. The character has {hair_color}. Keep the exact same hairstyle and structure as the input image. Masterpiece, cel shading, flat color."
\end{lstlisting}

\section{HairGPT Implementation Details}

\subsection{HairGPT Architecture Details}

The architecture of \textit{HairGPT} is designed to unify linguistic, visual, and geometric modalities into a single autoregressive framework. The following details describe the backbone, multimodal projectors, and the discrete strand tokenizer.

\paragraph{Transformer Backbone.}
We use the \textit{Qwen3-1.7B} decoder-only Transformer as the core generative backbone of HairGPT. The model operates on a hidden dimension of $D=2048$ and consists of standard self-attention layers and feed-forward networks. To accommodate hairstyle synthesis, the original linguistic vocabulary is extended into a unified vocabulary $\mathcal{V}$ that incorporates discrete geometric tokens via specific ID offsets.

\paragraph{Multimodal Condition Encoders.}
To bridge the gap between continuous visual/textual features and the discrete LLM space, we employ specialized encoders and projection layers:
\begin{itemize}
    \item \textbf{Vision Encoder:} We leverage a frozen \textit{DINOv3} (ViT-L/16) backbone to extract high-level visual features. For an input image $\mathcal{I} \in \mathbb{R}^{512 \times 512 \times 3}$, the encoder produces patch-level tokens. These are mapped to the Transformer's hidden dimension using a learnable lightweight MLP projector, $\mathcal{P}_{\text{img}}$, resulting in a sequence of visual embeddings $\mathbf{E}_{\text{img}}$.
    \item \textbf{Text Projector:} While we retain the pretrained word embeddings from Qwen, we apply a learnable linear projector $\mathcal{P}_{\text{txt}}$ to align the text features with the shared multimodal latent space, ensuring consistent conditioning across global and regional instructions.
\end{itemize}

\subsection{Training Details}
\paragraph{Cluster-based Token Augmentation.} 
To enhance the model's generalization and prevent it from overfitting to specific geometric instances, we implement a cluster-based strand token sampling strategy as a primary data augmentation technique. During training, instead of utilizing a fixed representative centroid for each of the $N_{\text{guide}}=512$ clusters, we dynamically sample a strand from the corresponding cluster pool (as described in Sec.~\ref{sec:sampling_strategy}). This stochastic selection ensures that the autoregressive transformer encounters varied geometric realizations of the same topological structure across different iterations. By introducing this instance-level variability, we effectively regularize the latent space, forcing the model to learn robust structural relationships and growth priors rather than memorizing individual strand coordinates.

\paragraph{Mode Selection Probability.}
To ensure the Transformer backbone $\Phi$ learns each hierarchical level of the hairstyle representation with equal proficiency, we adopt a mode-specific training scheme. During each training iteration, for every aligned triplet $(\mathcal{I}, \mathcal{T}, \mathcal{H})$, we randomly sample a generation mode to construct the input sequence $\mathbf{S}^{(\cdot)}$. We apply a uniform probability distribution for mode selection: $P(\text{Layout}) = 1/3$, $P(\text{Coarse}) = 1/3$, and $P(\text{Style}) = 1/3$. This balanced sampling prevents the model from overfitting to any single stage of the geometric synthesis and ensures a stable gradient flow across the entire generative hierarchy.

\paragraph{Multimodal Condition Dropout.}
To enhance the model's robustness to missing or noisy inputs, we implement a multimodal dropout strategy during training. Conditioning signals are randomly replaced with a learnable null embedding $\emptyset$ according to the following probabilities:
\begin{itemize}
    \item \textbf{Image Dropout:} The visual context $\mathbf{E}_{\text{img}}$ is dropped with a probability of $p_{\text{img}} = 0.3$.
    \item \textbf{Text Dropout:} The linguistic prompts $\mathbf{E}_{\text{txt}}$ (both global and regional) are dropped with a probability of $p_{\text{txt}} = 0.3$.
    \item \textbf{Joint Unconditional Training:} With a probability of $p_{\text{null}} = 0.1$, both image and text conditions are dropped simultaneously, forcing the model to perform unconditional generation based purely on the hairstyle distribution priors.
\end{itemize}
When a modality is dropped, its corresponding attention mask is set to zero to completely isolate the influence of the input. This strategy enables users to steer the generation process during inference by adjusting the guidance scale between conditional and unconditional logits.

\paragraph{Training Details and Hyperparameters.}
We train \textit{HairGPT} using 32 NVIDIA H20 GPUs for approximately 24 hours. The Transformer backbone is initialized from the pretrained \texttt{Qwen3-1.7B} weights to leverage its linguistic priors. Such initialization preserves pre-existing knowledge, enabling zero-shot understanding of fine-grained text attributes (e.g., ``curly'' and ``afro'').
For optimization, we employ the AdamW optimizer with a weight decay of 0.1. The learning rate follows a cosine decay schedule, starting from a base learning rate of $5 \times 10^{-5}$ and annealing to a final value of $1 \times 10^{-6}$. Notably, to facilitate rapid alignment between the geometric and multimodal latent spaces, we apply a $10\times$ learning rate multiplier ($5 \times 10^{-4}$) specifically to the vision and text projectors, while the Transformer backbone parameters are optimized using the base learning rate.

\subsection{Inference Details}

\label{sec:inference}

Hairstyle inference in \textit{HairGPT} is modeled as a progressive hierarchical unfolding process, transforming global multimodal conditions $\mathcal{C} = \{\mathcal{I}, \mathcal{T}\}$ into localized geometric details. Following the dual-decoupled parameterization, the synthesis is executed through a multi-pass generation flow to ensure rigorous structural control.

\paragraph{Phased Autoregressive Generation.} 
The inference process is partitioned into three distinct functional phases:
\begin{itemize}
    \item \textbf{Density Phase:} The model first generates a compressed sequence of density tokens $\mathbf{D}$. This establishes the macroscopic hair distribution and occupancy plan on the 2D scalp manifold.
    \item \textbf{Layout Phase:} Guided by the density prior and task separator $s_1$, the model autoregressively predicts spatial anchors $(u_k, v_k)$ for each of the $M=8$ semantic regions. This phase defines the precise location of every guide strand.
    \item \textbf{Coarse and Style Phases:} Once strand placements are fixed, the model enters the geometric refinement stages. Here, the transformer acts as a coordinate-conditioned mapper. By injecting the previously generated $(u_k, v_k)$ as a spatial prefix alongside mode separators $s_2$ or $s_3$, the model predicts the corresponding low-frequency backbones $T_{\text{coa}}$ and high-frequency residuals $T_{\text{sty}}$.
\end{itemize}
This "position-first, geometry-after" strategy ensures that complex geometric details are strictly grounded to the established physical layout, preventing spatial drift in long-sequence generation.

\paragraph{Task Steering via Separators.} 
To manage multiple synthesis objectives within a unified transformer backbone, we utilize explicit task separators $\{s_1, s_2, s_3\}$ as functional switches. In our implementation, these separators serve as tokens that steer the model's internal attention to focus on specific sub-tasks: $s_1$ for layout planning, $s_2$ for topological backbone synthesis, and $s_3$ for fine-grained textural detail. 

\paragraph{Compositional Editing.}
The decoupled nature of our inference process naturally enables flexible downstream applications. By strategically re-injecting prefixes and task separators, users can perform \textbf{compositional editing} without regenerating the entire hairstyle. For instance, by keeping the spatial anchors and coarse backbones constant and only re-sampling tokens following the $s_3$ separator, \textit{HairGPT} can perform style transfer or texture refinement (e.g., converting straight hair to coily hair) while preserving the original hair volume and global flow.